\documentclass[aip,jcp,amsmath,amssymb,reprint,superscriptaddress,floatfix]{revtex4-1}
\usepackage{graphicx,dcolumn,bm}
\usepackage[utf8]{inputenc}
\usepackage[T1]{fontenc}
\usepackage{mathptmx}
\usepackage{etoolbox}
\usepackage[unicode,hidelinks]{hyperref}
\hypersetup{
 pdftitle={Chiral-vibration-induced magnetism and circular dichroism in organic molecules},
 pdfauthor={Hao-Wen Xu, Cheng Xu, Zi-Yang Xu, Xiao-Long Yang, Hang Xie}}
\newlength{\JCPcolumnfigurewidth}
\newlength{\JCPwidefigurewidth}
\newcommand{\vect}[1]{\vec{#1}}
\newcommand{\Rea}{\operatorname{Re}}
\newcommand{\VCD}{\operatorname{VCD}_{E}}
\makeatletter
\def\@email#1#2{%
 \endgroup
 \patchcmd{\titleblock@produce}
  {\frontmatter@RRAPformat}
  {\frontmatter@RRAPformat{\produce@RRAP{*#1\href{mailto:#2}{#2}}}\frontmatter@RRAPformat}
  {}{}
}
\newenvironment{inplacetable}{%
 \par\smallskip\noindent\begin{minipage}{\linewidth}\def\@captype{table}\centering\footnotesize
}{\end{minipage}\par\smallskip}

\AtBeginDocument{\let\auto@bib@innerbib\@empty\let\write@bibliographystyle\relax}
\makeatother
\begin{document}

\raggedbottom
\setcounter{dbltopnumber}{2}

\setlength{\JCPcolumnfigurewidth}{\columnwidth}
\ifdim\JCPcolumnfigurewidth>85mm\setlength{\JCPcolumnfigurewidth}{85mm}\fi
\setlength{\JCPwidefigurewidth}{170mm}
\ifdim\JCPwidefigurewidth>\textwidth\setlength{\JCPwidefigurewidth}{\textwidth}\fi
\title{Chiral-vibration-induced magnetism and circular dichroism in organic molecules}
\author{Hao-Wen Xu}
\affiliation{College of Physics, Chongqing University, Chongqing 401331, China}
\author{Cheng Xu}
\affiliation{College of Physics, Chongqing University, Chongqing 401331, China}
\author{Zi-Yang Xu}
\affiliation{College of Physics, Chongqing University, Chongqing 401331, China}
\author{Xiao-Long Yang}
\affiliation{College of Physics, Chongqing University, Chongqing 401331, China}
\affiliation{Chongqing Key Laboratory for Strongly-Coupled Physics, Chongqing University, Chongqing 401331, China}
\author{Hang Xie\textsuperscript{*,}}
\email[Corresponding author: ]{xiehang2001@hotmail.com}
\affiliation{College of Physics, Chongqing University, Chongqing 401331, China}
\affiliation{Chongqing Key Laboratory for Strongly-Coupled Physics, Chongqing University, Chongqing 401331, China}
\begin{abstract}
We investigate the vibrational inverse Faraday effect and transient vibrational circular dichroism in benzene, coronene, and methane at the single-molecule level. We combine density-functional tight-binding calculations with damped normal-mode dynamics. Circularly polarized light excites suitable degenerate infrared-active modes, driving chiral nuclear motion that produces effective magnetic moments. Reversing the pump helicity reverses the vibrational chirality and the sign of the transient dichroic response. These results support a writing-readout scheme in which circularly polarized pumping prepares a chiral vibrational state and transient vibrational circular dichroism probes its handedness. We further examine how molecular symmetry shapes the light-induced chiral response across the three molecular systems. This work provides a valuable platform for exploring all-optical control of magnetism and chiroptical probing at the molecular scale.
\end{abstract}
\maketitle
\section{Introduction}\label{sec:introduction}
Chiral phonons carrying nonzero angular momentum have attracted considerable attention in recent years. Studies now span transport, nonlinear optics, and ultrafast dynamics, establishing chiral phonons as an emerging research field.\cite{r1,r2} One particularly promising direction is the interaction between chiral phonons and electronic orbital degrees of freedom. In conventional transition metals, orbital angular momentum can be generated through charge-to-orbital\cite{r3,r4} or spin-to-orbital\cite{r5,r6} conversion. Chiral-phonon--orbital coupling instead offers the possibility of generating orbital currents without a net charge current, providing a basis for low-power orbitronics. Chiral phonons can also possess appreciable effective magnetic moments, in some systems approaching the scale of the electronic Bohr magneton.\cite{r7,r8,r9} They may therefore mediate vibration-driven magnetism.

Circularly polarized excitation of suitable helicity can induce effective magnetic fields or magnetization through chiral phonons, an effect commonly termed the phonon inverse Faraday effect (phonon IFE). Related electric-field-driven dynamical multiferroicity has also been reported.\cite{r10} Recent work has addressed both its microscopic mechanism and its magnitude. Luo \textit{et al.}\cite{r11} observed large effective magnetic fields associated with chiral phonons in rare-earth halides. Shabala and Geilhufe\cite{r12} analyzed electron--phonon coupling pathways to clarify the dynamical origin of the phonon IFE and its connection to electronic structure. Lopez \textit{et al.}\cite{r13} subsequently considered a magneto-opto-phononic framework. The conditions for generating chiral phonons, and chiral vibrations more broadly, can be less restrictive than circularly polarized driving alone. Ren \textit{et al.}\cite{r14} showed theoretically that periodic linearly polarized excitation can produce coherent chiral phonons and nonzero steady-state magnetization in a paramagnet through nonlinear spin--phonon coupling and spontaneous symmetry breaking. The concept also extends beyond crystalline phonons: Feng \textit{et al.}\cite{r15} investigated vibrational angular momentum in thermally driven chiral molecular systems. These developments motivate the study of nonequilibrium chiral nuclear motion at the molecular scale.

Experimentally, chirality is manifested by different responses to left- and right-circularly polarized light in absorption, refraction, and scattering. Vibrational circular dichroism (VCD) measures the difference in mid-infrared absorption between the two helicities and is the vibrational counterpart of electronic circular dichroism.\cite{r16,r17} Following its initial observation in the 1970s,\cite{r18,r19} theoretical descriptions developed from classical treatments of optical polarization in molecules\cite{r20} to the non-Born--Oppenheimer electronic transition-current-density formulation of Nafie.\cite{r21} Polavarapu\cite{r22} reviewed quantum-mechanical calculations of VCD, vibrational Raman optical activity, and vibrational optical rotation. VCD has since become an established tool in physical and analytical chemistry.\cite{r16,r17} Advances in instrumentation and experimental sensitivity\cite{r23,r24} have broadened its use in studies of short-lived intermediates and nonequilibrium optical properties.\cite{r25,r26} Pollok \textit{et al.}\cite{r23} combined matrix isolation and VCD spectroscopy to identify the stereochemistry of conformational states. Dutta and Helbing\cite{r24} reported an actively stabilized interferometric arrangement for ultrafast chiral and achiral infrared spectroscopy. Studies by Sato \textit{et al.}\cite{r25} and Taniguchi \textit{et al.}\cite{r26} further examined intermediate states in gelation and the evolution of helical supramolecular systems. Together, these studies motivate time-resolved optical probes of molecular and supramolecular chirality.

A microscopic dynamical description connecting chiral molecular vibrations to magnetism remains an important objective. Most phonon-IFE studies have focused on crystalline condensed-matter systems,\cite{r11,r12,r13} where the lattice environment complicates the relation between nuclear motion and magnetic response. Here, we develop a molecular framework for the vibrational inverse Faraday effect (vibrational IFE), using isolated organic molecules to resolve the buildup and evolution of vibration-induced magnetism. Molecular point groups, including $D_{6h}$, $T_d$, and more broadly $C_5$, provide a range of symmetry settings for examining degeneracy and chiral motion. A molecular description also permits atom-resolved analysis without the additional complexity of a crystal lattice. We investigate an optical writing-readout scheme: circularly polarized light (CPL) prepares one of two vibrational states of opposite chirality in an initially achiral molecule, and transient VCD probes the resulting handedness. In the proposed scheme, the pump helicity, the direction of vibrational rotation, and the sign of the VCD response are directly related. In addition, the early oscillatory buildup of the magnetic moment provides a time-domain feature that can distinguish it from the initially linear growth of electronic orbital magnetization reported in Ref.~\onlinecite{r27}.

\section{Theory}\label{sec:theory}
\subsection{Equations of motion}\label{sec:eom}
Consider a molecule containing $N$ atoms with $n=3N$ Cartesian degrees of freedom. The atomic displacement vector is used as the generalized coordinate, $\vect q=(q_1,q_2,\ldots,q_n)^{\mathrm T}$. Near the equilibrium geometry, the potential energy is approximated by
\begin{equation}\label{eq:potential}
 V(\vect q)\approx \frac12\vect q^{\mathrm T}\mathbf H\vect q,
\end{equation}
where $\mathbf H$ is the $n\times n$ Hessian with elements $H_{ij}=(\partial^2 V/\partial q_i\partial q_j)_{\vect q=\vect0}$. The kinetic energy is
\begin{equation}\label{eq:kinetic}
 T(\dot{\vect q})=\frac12\dot{\vect q}^{\mathrm T}\mathbf M\dot{\vect q},
\end{equation}
where $\mathbf M$ is the diagonal mass matrix. Substituting $L=T-V$ into the Euler--Lagrange equations,
\begin{equation}\label{eq:euler}
 \frac{d}{dt}\left(\frac{\partial L}{\partial\dot q_i}\right)-\frac{\partial L}{\partial q_i}=0,
 \qquad i=1,\ldots,n,
\end{equation}
gives the linear equations of motion,
\begin{equation}\label{eq:free}
 \mathbf M\ddot{\vect q}+\mathbf H\vect q=\vect0.
\end{equation}
To describe vibrational relaxation,\cite{r28} we introduce a linear damping term proportional to velocity. The mode-dependent damping models energy loss through van der Waals coupling to resonant vibrations of surrounding molecules in an effective fluid environment. We use a symmetric damping matrix $\mathbf C$ constructed to be diagonal in the mass-weighted normal-mode basis. Its modal eigenvalues are nonnegative, with positive rates for the damped vibrational modes. The equations become
\begin{equation}\label{eq:damped}
 \mathbf M\ddot{\vect q}+\mathbf C\dot{\vect q}+\mathbf H\vect q=\vect0.
\end{equation}

\subsection{Normal-mode analysis and decoupling}\label{sec:normal}
Introduce mass-weighted coordinates $\vect\xi=\mathbf M^{1/2}\vect q$ and left-multiply both sides of Eq.~\eqref{eq:damped} by $\mathbf M^{-1/2}$. This congruence transformation preserves the real symmetry of the relevant matrices:
\begin{equation}\label{eq:weighted}
 \ddot{\vect\xi}+\widetilde{\mathbf C}\dot{\vect\xi}+\mathbf D\vect\xi=\vect0,
\end{equation}
with $\widetilde{\mathbf C}=\mathbf M^{-1/2}\mathbf C\mathbf M^{-1/2}$ and $\mathbf D=\mathbf M^{-1/2}\mathbf H\mathbf M^{-1/2}$. Solving $\mathbf D\mathbf U=\mathbf U\boldsymbol\Lambda$ gives the orthonormal eigenvector matrix $\mathbf U$, satisfying $\mathbf U^{\mathrm T}\mathbf U=\mathbf I$, and the diagonal matrix $\boldsymbol\Lambda=\mathbf U^{\mathrm T}\mathbf D\mathbf U=\operatorname{diag}(\omega_1^2,\ldots,\omega_n^2)$. The $\omega_k$ are the normal-mode angular frequencies. The normal coordinates are
\begin{equation}\label{eq:normalcoordinates}
 \vect Q=\mathbf U^{\mathrm T}\vect\xi.
\end{equation}
Substituting $\vect\xi=\mathbf U\vect Q$ into Eq.~\eqref{eq:weighted} and left-multiplying both sides by $\mathbf U^{\mathrm T}$ yields
\begin{equation}\label{eq:modal}
 \ddot{\vect Q}+\boldsymbol\Gamma\dot{\vect Q}
 +\boldsymbol\Lambda\vect Q=\vect0.
\end{equation}
Here the diagonal modal damping matrix is
\[
\begin{aligned}
 \boldsymbol\Gamma
 &=\mathbf U^{\mathrm T}\widetilde{\mathbf C}\mathbf U\\
 &=\operatorname{diag}(\gamma_1,\gamma_2,\ldots,\gamma_{3N}).
\end{aligned}
\]
Each coefficient $\gamma_k$ describes energy dissipation in the corresponding normal mode. The system therefore separates into $n=3N$ damped harmonic oscillators:
\begin{equation}\label{eq:modefree}
 \ddot Q_k+\gamma_k\dot Q_k+\omega_k^2Q_k=0,\qquad k=1,\ldots,n.
\end{equation}
Here we note that repeated eigenvalues correspond to degenerate vibrational modes. Their degeneracy is constrained by molecular symmetry, as discussed in Appendix~\ref{app:group}, and provides a basis for the chiral motions considered here.

\subsection{Response to a time-harmonic electric field}\label{sec:driven}
Including the force of an external time-harmonic electric field gives
\begin{equation}\label{eq:driven}
 \mathbf M\ddot{\vect q}+\mathbf C\dot{\vect q}+\mathbf H\vect q=\vect F.
\end{equation}
The corresponding mass-weighted equation is
\begin{equation}\label{eq:weighteddriven}
 \ddot{\vect\xi}+\widetilde{\mathbf C}\dot{\vect\xi}+\mathbf D\vect\xi=\widetilde{\vect F}.
\end{equation}
Here, we define $\widetilde{\vect F}=\mathbf M^{-1/2}\vect F$. Transforming to normal coordinates gives
\begin{equation}\label{eq:normaldriven}
 \ddot{\vect Q}+\boldsymbol\Gamma\dot{\vect Q}
 +\boldsymbol\Lambda\vect Q=\vect G,
\end{equation}
where $\vect G=\mathbf U^{\mathrm T}\widetilde{\vect F}$. The $k$th component of Eq.~\eqref{eq:normaldriven} is
\begin{equation}\label{eq:modedriven}
 \ddot Q_k+\gamma_k\dot Q_k+\omega_k^2Q_k=G_k(t).
\end{equation}
Each coordinate $Q_k$ thus describes a damped normal mode with damping coefficient $\gamma_k$, natural angular frequency $\omega_k$, and driving force $G_k(t)$. For a harmonic force $G_k(t)=G_k\cos\omega t$ in the weak-damping regime, Eq.~\eqref{eq:modedriven} has the solution
\begin{equation}\label{eq:solution}
\begin{aligned}
 Q_k(t)={}&Q_{0k}e^{-\gamma_k t/2}\cos(\omega'_k t+\phi_k)\\
 &+\frac{G_k\bigl[(\omega_k^2-\omega^2)\cos\omega t
          +\gamma_k\omega\sin\omega t\bigr]}
 { (\omega_k^2-\omega^2)^2+\gamma_k^2\omega^2}.
\end{aligned}
\end{equation}
Here $Q_{0k}$ and $\phi_k$ are integration constants fixed jointly by the initial displacement and velocity after accounting for the driven particular solution. In general, $Q_{0k}$ is not equal to $Q_k(0)$. The damped angular frequency is $\omega'_k=\sqrt{\omega_k^2-\gamma_k^2/4}$, and $G_k$ is the modal force amplitude. All molecular calculations considered here are in the weak-damping regime, $\gamma_k/\omega_k\ll1$ for the resonant vibrational modes. General small-amplitude motion is a linear superposition of these mode responses.

The linear absorption spectrum is obtained by adding the steady-state contributions of the normal modes. In the weak-damping and near-resonance limits, $\gamma_k\ll\omega_k$ and $\omega\approx\omega_k$, we adopt the Lorentzian form
\begin{equation}\label{eq:lorentz}
 I_k(\omega)=\frac{A_k(\gamma_k/2)}{(\omega-\omega_k)^2+(\gamma_k/2)^2},
\end{equation}
Here $A_k$ is a spectral-weight factor incorporating the chosen intensity normalization. With $\gamma_k$ defined as the coefficient of $\dot Q_k$, the near-resonant full width at half maximum (FWHM) in angular frequency is $\gamma_k$, and the half width is $\gamma_k/2$. The FWHM in ordinary frequency is $\gamma_k/(2\pi)$. Degenerate-mode intensities are added. Retaining the full Cartesian space, the total response is
\begin{equation}\label{eq:spectrum}
 I(\omega)=\sum_{k=1}^{3N}I_k(\omega).
\end{equation}
The calculation retains all $3N$ Cartesian coordinates, including rigid-body motion. The near-resonant form in Eq.~\eqref{eq:lorentz} is used only for positive-frequency vibrational resonances; rigid-body responses are evaluated from the full equations of motion.

The energetic observable underlying the relative spectra is the mean elastic potential energy,
\begin{equation}\label{eq:observable}
 \overline E_{\mathrm{el}}(\omega)
 =\frac{1}{2\tau}\int_0^\tau\vect q^{\mathrm T}(t;\omega)\mathbf H\vect q(t;\omega)\,dt.
\end{equation}
Under illumination, absorbed optical energy is converted into elastic potential energy and kinetic energy of molecular vibrations, and is subsequently dissipated as heat. In the single-mode, weak-damping, near-resonant steady state, optical absorption balances dissipation, so the mean absorbed power is proportional to the mean elastic energy: $\overline P_k\propto\gamma_k\overline E_{\mathrm{el},k}$, with $\overline P_k\approx2\gamma_k\overline E_{\mathrm{el},k}$ under the damping convention used here. This relation justifies using the elastic-energy spectrum to represent relative absorption.

\subsection{Chiral vibration and the vibrational-IFE magnetic moment}\label{sec:magnetism}
Just as circular polarization is formed by phase-shifted oscillations in two orthogonal directions, two molecular vibrational components with the same frequency and suitable orthogonal displacement directions can combine into an elliptical or circular trajectory. Degenerate orthogonal vibrational components can therefore support chiral motion when excited with the appropriate relative phase. We represent this superposition by
\begin{equation}\label{eq:chiral}
 \vect Q_{\mathrm{chiral}}(t)=\operatorname{Re}\!\left[
 \vect\alpha e^{i\omega t}+\vect\beta e^{i(\omega t+\pi/2)}\right].
\end{equation}
Here $\vect\alpha$ and $\vect\beta$ are real, nonzero, orthogonal amplitude vectors in normal-coordinate space. Their magnitudes need not be equal: equal magnitudes give a circle in this space, whereas unequal magnitudes give an ellipse. Taking the real part gives the real normal coordinates, and the sign of the relative phase fixes the rotation direction. Physical atomic displacements follow from $\vect q=\mathbf M^{-1/2}\mathbf U\vect Q_{\mathrm{chiral}}$. Orthogonality in normal-coordinate space does not require orthogonal displacement components at every atom; a local chiral trajectory requires the two displacement components at that atom to be non-collinear.
The proposed criterion is examined below for molecules with different point-group symmetries.

The circulating motion of charged atoms produces a magnetic moment. The gyromagnetic relation is written as
\begin{equation}\label{eq:atomicmu}
 \vect\mu=\frac{q}{2m}\vect L=\frac q2\vect r\times\vect v,
\end{equation}
Here $\vect L$, $\vect v$, and $\vect r$ are the atomic angular momentum, velocity, and position relative to the chosen origin, respectively; $q$ is the scalar effective atomic charge. In the magnetic-moment expressions, $q$ or $q_i$ denotes charge, distinct from the displacement vector $\vect q$ and its Cartesian components used above. Summing over atoms gives
\begin{equation}\label{eq:totalmu}
 \vect\mu_{\mathrm{total}}=\sum_{i=1}^{N}\frac{q_i}{2}\vect r_i\times\vect v_i.
\end{equation}
These expressions specify our calculation of the light-induced magnetic moment. The values of $\vect L$ and $\vect r$ depend on the origin; the molecular center of mass is used here. Because of the symmetry of the molecular structures considered, we find in our calculations that summing moments referred to individual atomic equilibrium positions gives the same total result as using the molecular center of mass.

\subsection{Transient vibrational circular dichroism}\label{sec:vcd}
To probe the chirality prepared by CPL, the molecule is first driven to a fully established vibrational steady state. At $t=0$, the pump is switched off and left- and right-handed CPL probes are applied in separate cases. Atomic positions and velocities are inherited directly from the end of the pump stage. The probe-to-pump electric-field amplitude ratio is $E_{\mathrm{probe}}/E_{\mathrm{pump}}=0.01$. The pump and probe fields are in phase at switching. We define an energy-based transient dichroic signal,
\begin{equation}\label{eq:vcd}
 \operatorname{VCD}_{E}(\omega)
 =\overline E_{\mathrm{el},\mathrm L}(\omega)
 -\overline E_{\mathrm{el},\mathrm R}(\omega),
\end{equation}
The overbar denotes an average over the same probe interval $[0,\tau]$. At each scanned frequency, the probe window spans 100 probe periods, $\tau(\omega)=200\pi/\omega$. The signal represents the difference in relative absorption response between the two probe helicities through the mean elastic-energy difference; it is not an independently calculated transmission or absorbed-power difference. For the equal-stiffness pair in Appendix~\ref{app:vcd}, it equals $(k/2)$ times the corresponding mean-square-displacement difference. The energy-normalized signal $\operatorname{VCD}_{x^2}=(2/k)\operatorname{VCD}_{E}$ therefore equals that displacement difference directly.

\subsection{DFTB+ and computational details}\label{sec:computation}
Geometry optimization, electronic-structure calculations, and Hessian matrices were obtained with the density-functional tight-binding (DFTB) package DFTB+, version 22.2.\cite{r29,r30} DFTB is a semiempirical approximation to density-functional theory. The method is based on a second-order expansion of the total-energy expression, reducing computational cost while retaining a quantum-mechanical description suitable for molecular vibrational calculations.

The computational procedure comprises the following steps. First, the initial benzene structure is constructed as a planar regular hexagon with a C--C distance of $0.140\,\mathrm{nm}$ and a C--H distance of $0.108\,\mathrm{nm}$. Second, the geometries are optimized using the \texttt{LBFGS} algorithm until all atomic force components are below $1\times10^{-6}$ atomic units and the energy change is below $1\times10^{-14}$ atomic units. The \texttt{3ob} parameter set is used for the relevant organic elements, including C and H. Third, the Hessian is obtained at the optimized geometry using \texttt{SecondDerivatives}. The resulting force-constant matrix is used for the normal-mode analysis.
The damping provides an effective description of intermolecular vibrational relaxation; microscopic anharmonic relaxation mechanisms are discussed in Ref.~\onlinecite{r28}. We use it to represent a fluid environment without explicitly propagating the surrounding molecules. The calculations for benzene, coronene, and methane use the same damping parameter. Comparisons of the molecular responses below therefore do not involve changes in the damping parameter.
This choice of method balances computational efficiency and molecular description across the systems examined. The electric-field amplitude used in the optical calculations is $10^7\,\mathrm{V\,m^{-1}}$, which can be reached with ultrashort laser fields.

\section{Results and discussion}\label{sec:results}
Here, we examine the vibrational levels, infrared spectra, degenerate modes, and time-dependent magnetic responses of the three organic molecules. We then analyze the transient VCD following CPL excitation and use group theory to relate the observed responses to molecular symmetry.

\subsection{Benzene}\label{sec:benzene}
\begin{figure*}[tp]
 \centering
 \includegraphics[width=\JCPwidefigurewidth]{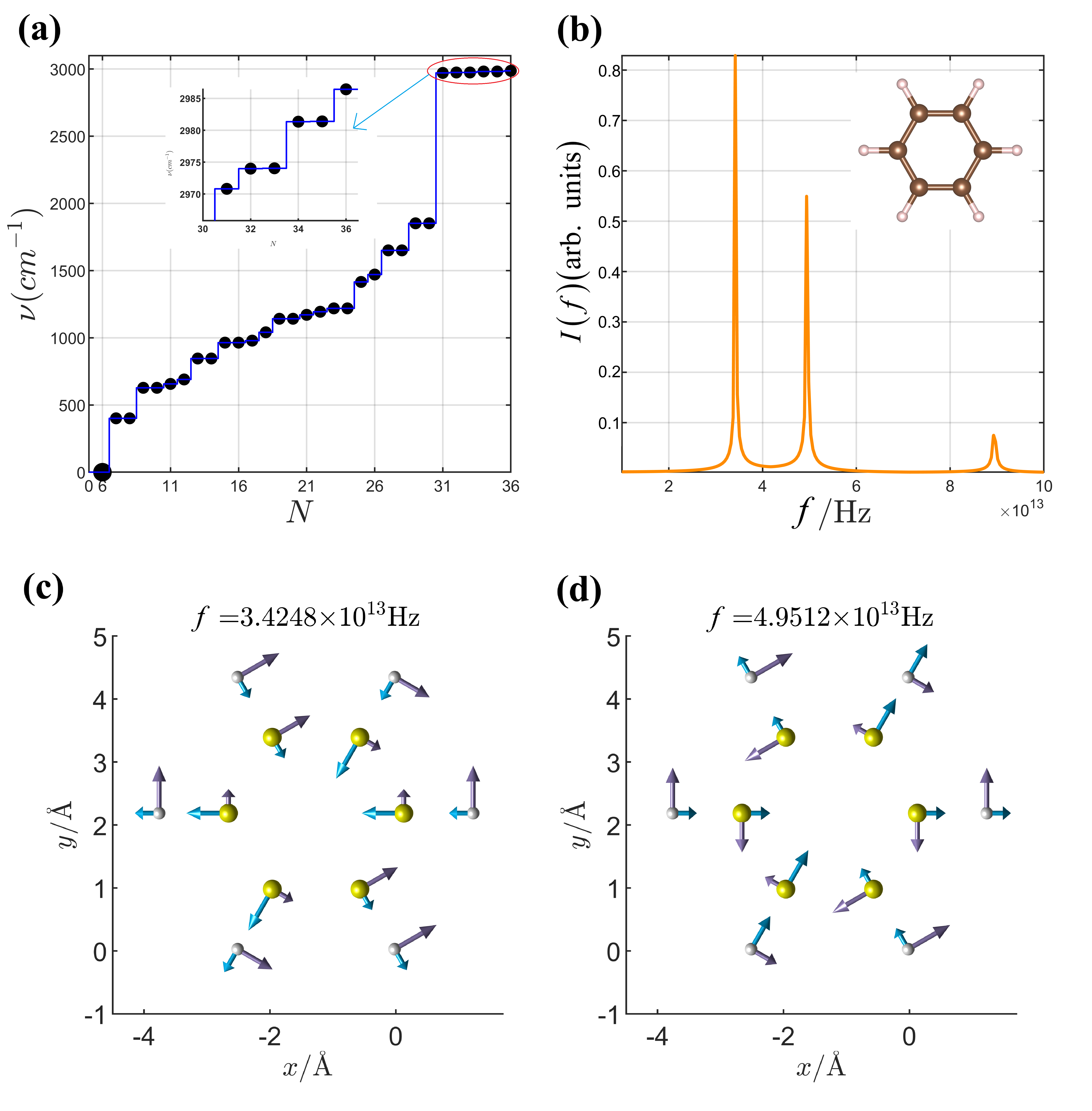}
 \caption{(a) Vibrational levels of benzene. (b) Infrared absorption spectrum; inset: molecular structure. (c) and (d) The two degenerate vibrational pairs discussed as mode A and mode B, respectively.}
 \label{fig:1}
\end{figure*}

\begin{figure*}[tp]\centering
\begin{minipage}[t]{\JCPcolumnfigurewidth}\vspace{0pt}
 \centering
 \includegraphics[width=\JCPcolumnfigurewidth]{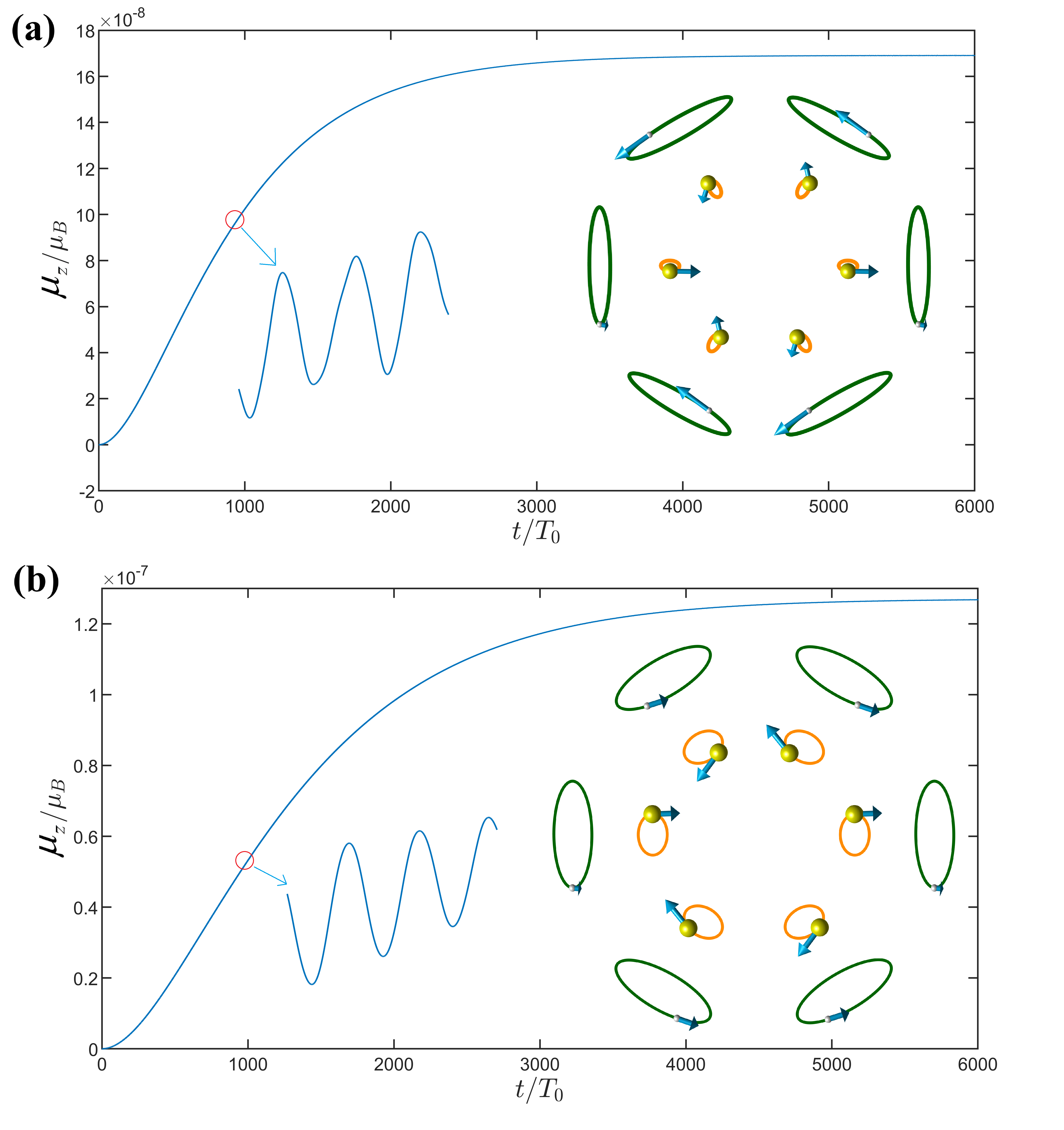}
 \caption{Magnetic-moment evolution under left-handed CPL excitation at the respective resonant frequencies of the benzene modes in Figs.~\ref{fig:1}(c) and \ref{fig:1}(d). (a) Mode A. (b) Mode B. Insets show the corresponding steady-state atomic trajectories over one vibrational period.}
 \label{fig:2}
\end{minipage}\hspace{2mm}\begin{minipage}[t]{\JCPcolumnfigurewidth}\vspace{0pt}
 \centering
 \includegraphics[width=\JCPcolumnfigurewidth]{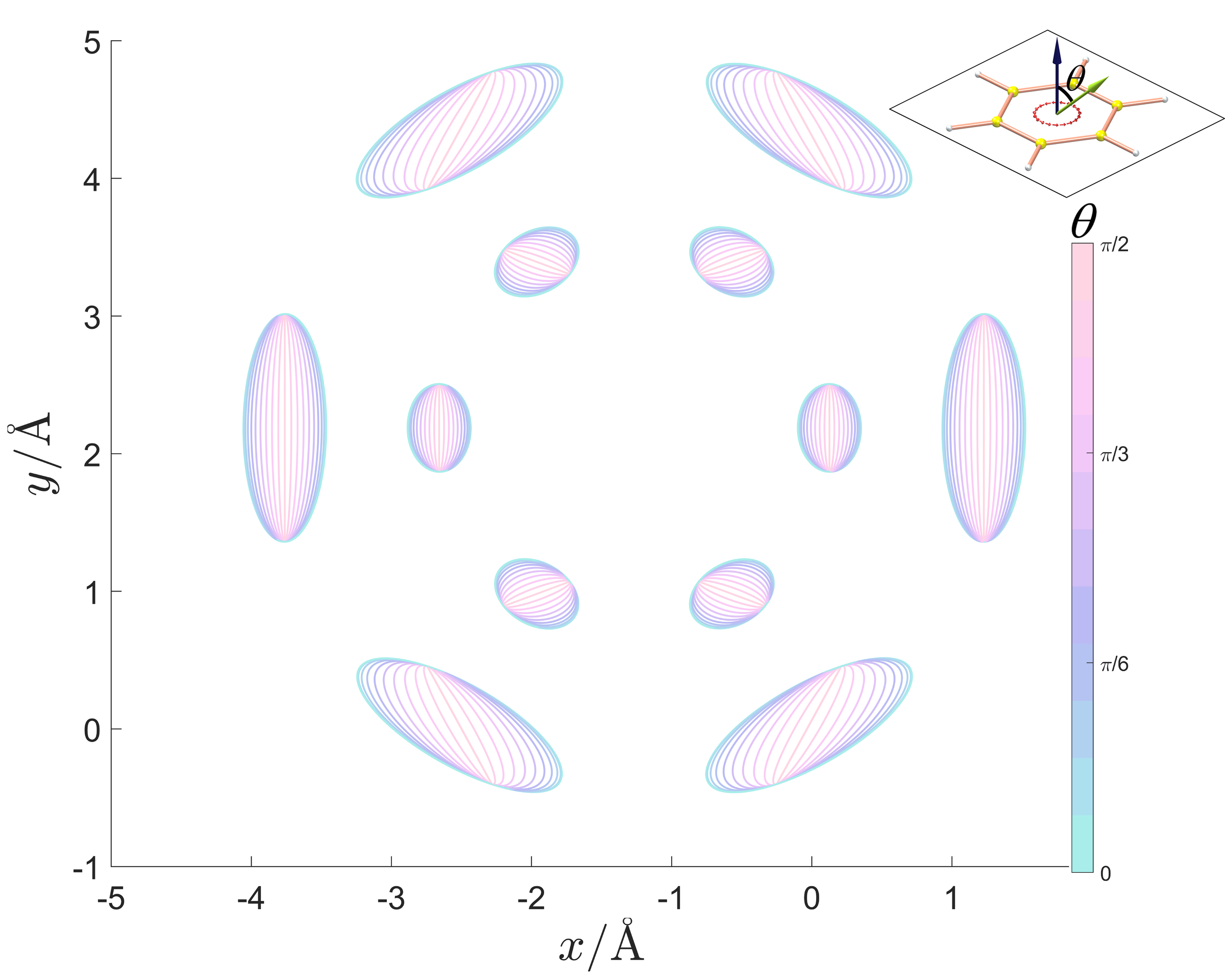}
 \caption{Chiral vibrational trajectories of benzene under CPL excitation at different incidence angles for mode B. The angle is measured from the normal to the molecular plane.}
 \label{fig:3}
\end{minipage}
\end{figure*}

Figure~\ref{fig:1}(a) shows the vibrational levels of benzene calculated from the Hessian obtained using the DFTB+ software. The molecule has 30 internal vibrational degrees of freedom, classified by the irreducible representations of $D_{6h}$. Doubly degenerate modes of symmetry types such as $E_{1u}$ and $E_{2g}$ appear as paired levels. The high-frequency levels near $90\,\mathrm{THz}$, indexed 31--36 when rigid translations and rotations are included, correspond to C--H stretching. Their infrared-active component produces the peak near $89.3\,\mathrm{THz}$ in Fig.~\ref{fig:1}(b). All three absorption peaks shown belong to doubly degenerate $E_{1u}$ channels. The calculated C--H stretching frequencies can be compared with the experimental frequencies in Ref.~\onlinecite{r31}.
The two lower-frequency absorption peaks occur near $34.2$ and $49.5\,\mathrm{THz}$. Figures~\ref{fig:1}(c) and \ref{fig:1}(d) illustrate the corresponding degenerate pairs, denoted mode A and mode B. Owing to $D_{6h}$ symmetry, equivalent atoms have the same vibration amplitude and symmetry-related displacement directions. Combining the two components of each pair produces collective elliptical atomic motions, as shown in Fig.~\ref{fig:2}. By contrast, the high-frequency C--H stretching pair has nearly collinear atomic displacements along the bond axes. Its superposition remains approximately linearly polarized and does not produce the chiral trajectories considered here.

Figure~\ref{fig:2} shows the time evolution of the magnetic moment under resonant continuous-wave CPL. The molecule initially occupies its equilibrium geometry with zero magnetic moment. For mode A at $34.2\,\mathrm{THz}$, left-handed CPL produces a moment that rises from zero and approaches saturation after approximately $3000T_0$. The saturation value depends on the field strength, effective atomic charges, and damping. At early times, $t<1500T_0$, the growth is accompanied by rapid oscillations at the driving frequency. We attribute this behavior to the different responses of the two vibrational components to CPL. Atomic trajectories initially spiral outward as their amplitudes grow and close into steady ellipses upon saturation. Reversing the pump helicity reverses the magnetic moment while preserving its magnitude.

The insets show the elliptical or nearly circular trajectories of the C and H atoms. Both modes are excited by left-handed CPL at their respective resonant frequencies. In mode A, the C and H trajectories rotate in the same direction; in mode B, they rotate in opposite directions. Such nuclear motion corresponds classically to circulating ionic currents and produces a magnetic moment perpendicular to the molecular plane. Because carbon carries a negative effective charge, opposite to that of hydrogen, its magnetic moment points opposite to its angular momentum.

Mode B at $49.5\,\mathrm{THz}$ has a higher frequency and a different pattern of C--H rotation, resulting in a different saturation moment. The magnetic response depends on frequency, effective charge, and the area enclosed by each atomic orbit. For a closed elliptical trajectory, the current-loop relation is $\mu=IS$, with signed current $I=qf$ and area $S=\pi ab$. Hence,
\begin{equation}\label{eq:ellipse}
 \mu=q\pi f ab=\frac q2\omega ab,
\end{equation}
where $a,b>0$ are the semiaxes, $f=\omega/(2\pi)>0$, and $q$ is the signed atomic charge. The scalar $\mu$ denotes the component along the oriented normal $\hat{\vect n}$ fixed by $\vect r\times\vect v$. Reversing the orbit reverses this normal, consistently with Eq.~\eqref{eq:atomicmu}. The magnetic-moment magnitude is $|q|\omega ab/2$. The total magnetic moment is separated into hydrogen and carbon contributions:
\begin{equation}\label{eq:CH}
 \mu_{\mathrm{total}}=\mu_{\mathrm H}+\mu_{\mathrm C}.
\end{equation}
The two contributions have opposite signs for mode A and the same sign for mode B. Although mode B benefits from its higher frequency, counter-rotation of C and H, and more circular trajectories, mode A absorbs much more strongly at resonance [Fig.~\ref{fig:1}(b)] and produces the larger total moment. The circularity of an orbit, related to its eccentricity $e$, also characterizes its vibrational chirality. As $e$ approaches unity, an ellipse approaches a line and its magnetic contribution vanishes.

These results connect CPL excitation, phase-coherent degenerate nuclear motion, circulating atomic charges, and an effective molecular magnetic moment. They extend the physical picture of the phonon IFE\cite{r11,r12} to an isolated organic molecule. Even an achiral equilibrium structure can acquire a net magnetic response through selective excitation of degenerate infrared-active modes. Its magnitude depends on the displacement geometry, oscillator strength, and field parameters.

Figure~\ref{fig:3} shows the steady-state atomic trajectories of mode B as the angle $\theta$ between the light wave vector and the molecular-plane normal varies from $0^\circ$ to $90^\circ$. Each closed curve represents one vibrational period. The normal-incidence trajectories are elliptical rather than perfectly circular. In CPL, the equal-amplitude, phase-shifted in-plane field components excite the orthogonal partners. The resulting normal-coordinate motion is then reconstructed in the local Cartesian coordinates of each atom. Although the collective mode amplitudes are equal in the symmetric driving limit, their local eigenvector components need not have equal lengths or be mutually orthogonal. The reconstructed atomic motion can therefore have unequal principal-axis amplitudes and trace an ellipse.
For the incidence plane chosen in the calculation, the in-plane $y$ component of the field decreases as $\cos\theta$, while the $x$ component remains unchanged. The resulting ellipses therefore change eccentricity and orientation rather than shrinking uniformly. At grazing incidence, $\theta\to90^\circ$, the in-plane field becomes linearly polarized, the closed orbit collapses to a line, and vibrational chirality is suppressed. Incidence angle thus offers a means of changing the vibration amplitude and induced moment at fixed field strength and frequency. Normal incidence maximizes the in-plane chiral driving in this geometry and is used for the pump--probe response below.

\begin{figure*}[tp]
 \centering
 \includegraphics[width=\JCPwidefigurewidth]{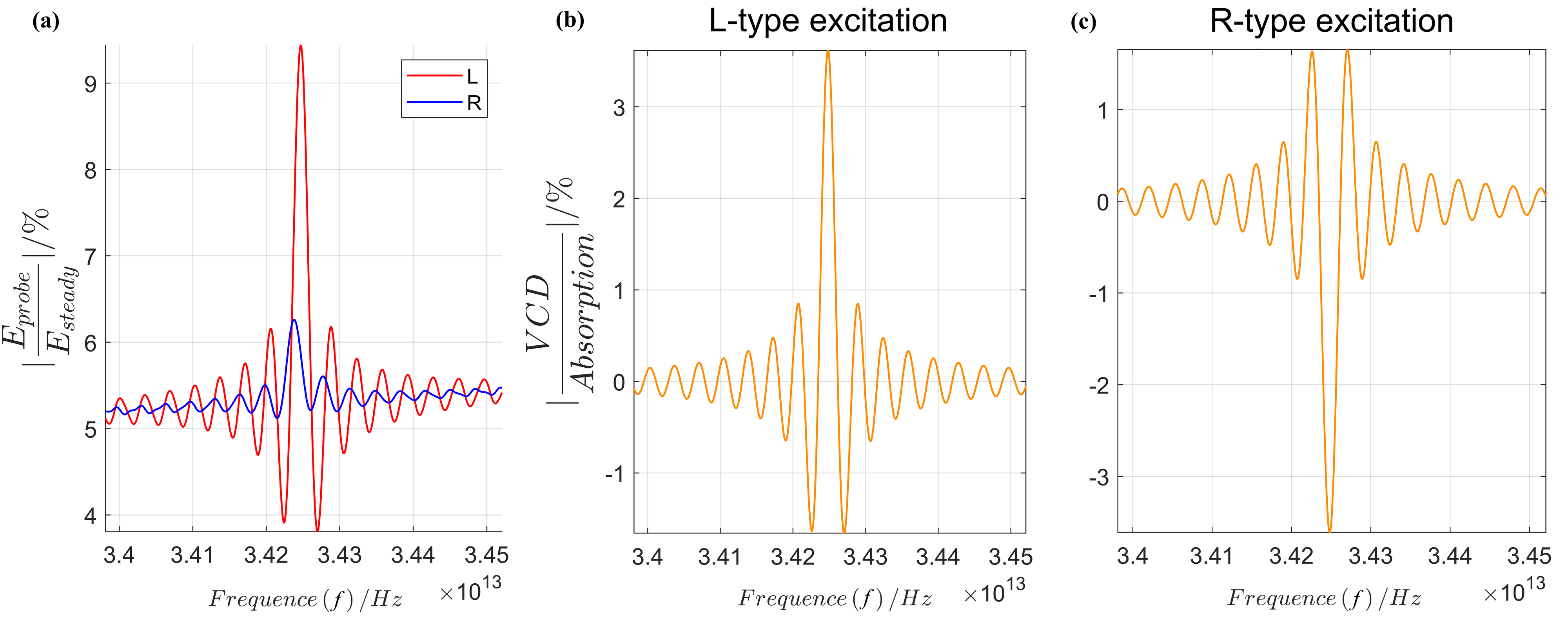}
 \caption{Excitation and detection of chirality-dependent vibrational responses in benzene. (a) Relative spectra for left- and right-handed probe CPL following steady-state preparation by left-handed CPL. (b) and (c) Transient VCD under resonant pumping with left- and right-handed CPL, respectively.}
 \label{fig:4}
\end{figure*}

Having established optical preparation of rotating nuclear motion, Fig.~\ref{fig:4} examines its readout through transient VCD. The molecule is pumped by CPL to a vibrational steady state and then probed with the two helicities in separate calculations. After left-handed pumping, the left-handed probe response exceeds the right-handed response, most prominently near mode A at $34.2\,\mathrm{THz}$, and the difference decreases away from resonance. Since equilibrium benzene is achiral, this difference is assigned to the prepared vibrational state.
Figures~\ref{fig:4}(b) and \ref{fig:4}(c) give the absorption difference between the relative responses to left- and right-handed probes, represented here by the mean elastic-energy difference defined in Eq.~\eqref{eq:vcd}. The response is sinc-like, with an extremum near resonance and decaying oscillatory side lobes. Changing the pump from left- to right-handed reverses the spectrum at every frequency. The two pump helicities therefore prepare oppositely rotating vibrational states whose handedness is encoded in the sign of the transient response. Figures~\ref{fig:2}--\ref{fig:4} establish the proposed optical writing-readout sequence within vibrational degrees of freedom, without invoking electronic transitions. This motivates examination of more complex organic molecules.

\subsection{Coronene}\label{sec:coronene}
\begin{figure*}[tp]
 \centering
 \includegraphics[width=\JCPwidefigurewidth]{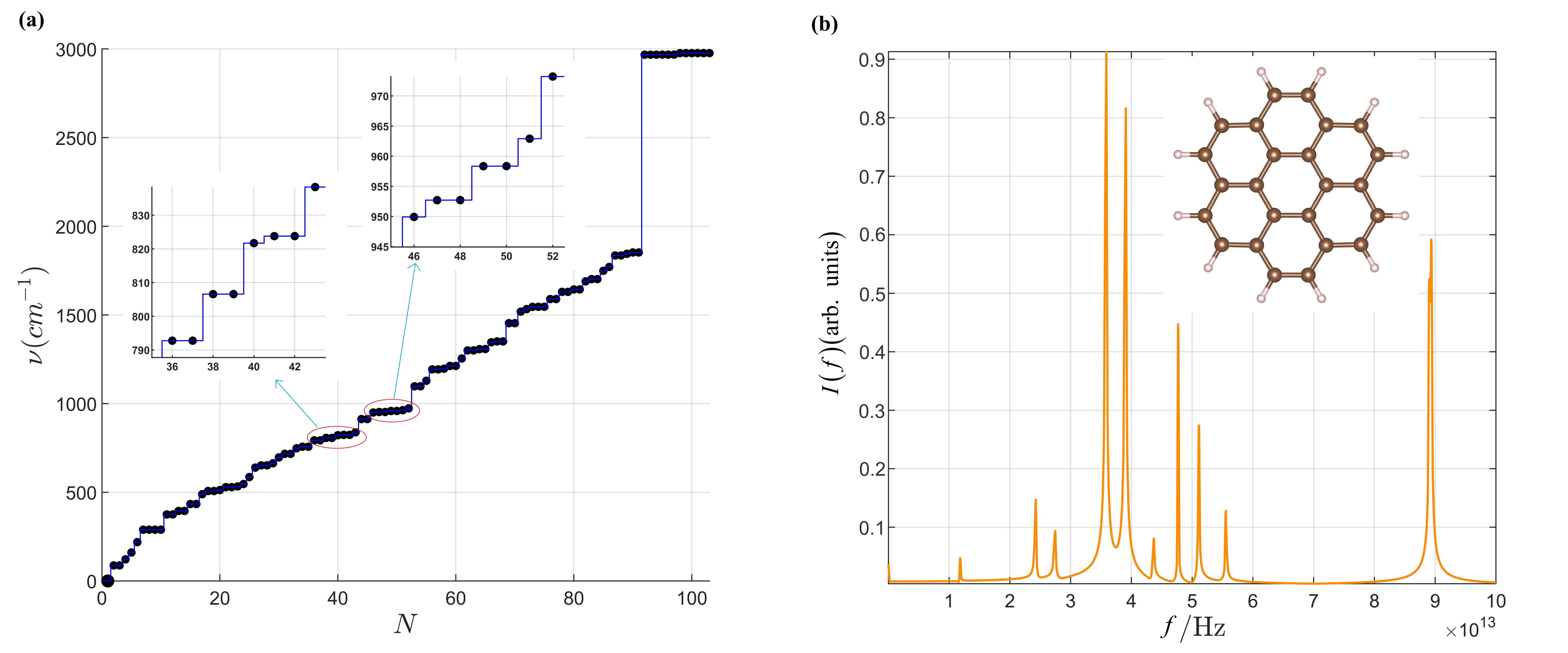}
 \caption{(a) Vibrational levels of coronene; the two insets magnify the circled energy-level regions. (b) Infrared absorption spectrum; inset: molecular structure.}
 \label{fig:5}
\end{figure*}

Coronene, $\mathrm{C}_{24}\mathrm{H}_{12}$, consists of a central benzene ring surrounded by six fused rings. Its planar equilibrium structure belongs to $D_{6h}$ [inset of Fig.~\ref{fig:5}(b)]. The Cartesian Hessian has 108 eigenvectors, of which 102 correspond to internal vibrations after removing rigid translations and rotations. Compared with benzene, the larger framework has a much higher mode density. Doubly degenerate $E$ representations, including $E_{1u}$ and $E_{2g}$, give paired levels. The high-frequency group near $90\,\mathrm{THz}$ (approximately $3000\,\mathrm{cm^{-1}}$) is associated with the 12 C--H stretches, whose infrared-active components belong to $E_{1u}$.
The spectrum in Fig.~\ref{fig:5}(b) is constructed from the damped normal-mode contributions in Eq.~\eqref{eq:spectrum}. The two principal peaks at $35.8$ and $39.0\,\mathrm{THz}$ correspond to $E_{1u}$ pairs, denoted mode A and mode B. The shift from the lower benzene resonance at $34.2\,\mathrm{THz}$ is associated with the greater rigidity of the fused-ring framework.

\begin{figure*}[tp]
 \centering
 \includegraphics[width=\JCPwidefigurewidth]{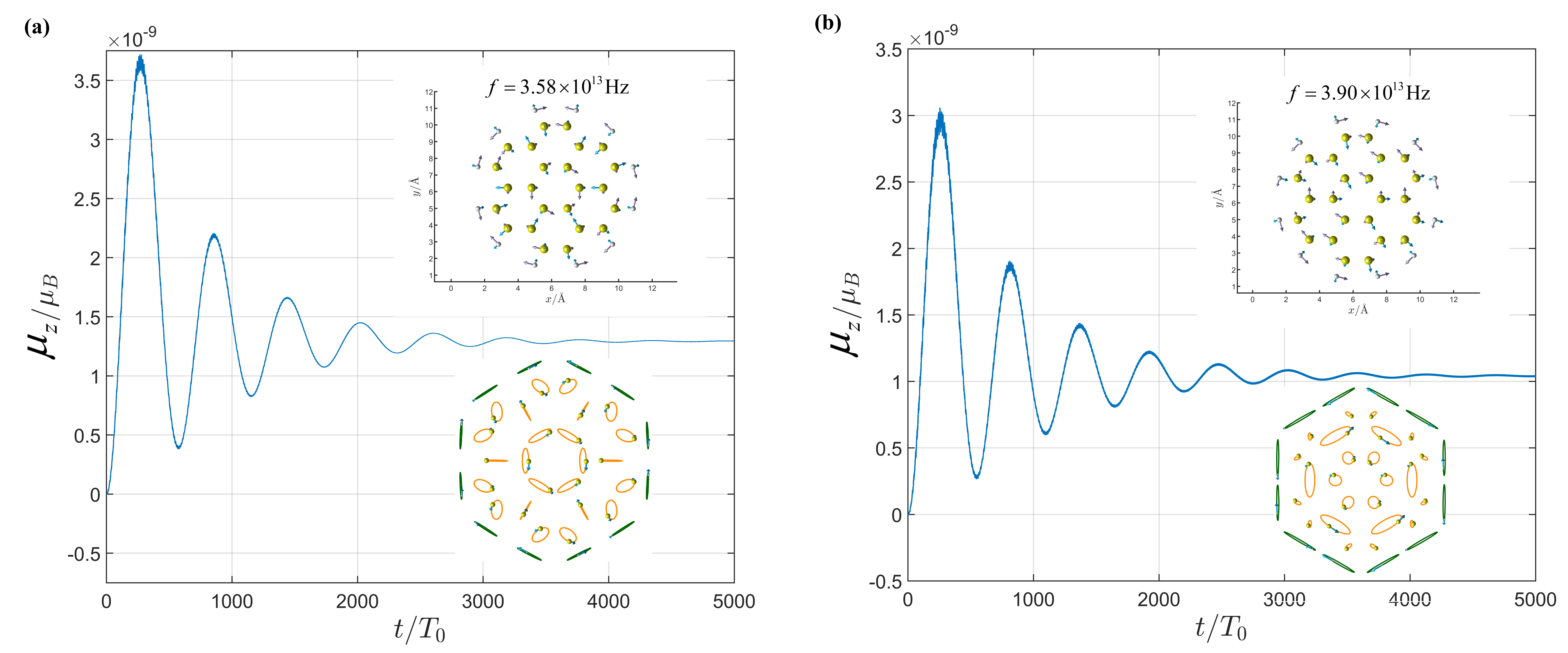}
 \caption{Magnetic-moment evolution under CPL excitation at the respective resonant frequencies of the coronene modes identified in Fig.~\ref{fig:5}. (a) Mode A at $35.8\,\mathrm{THz}$. (b) Mode B at $39.0\,\mathrm{THz}$. Upper-right insets: degenerate normal modes within each $E_{1u}$ pair. Lower-right insets: real-space atomic trajectories.}
 \label{fig:6}
\end{figure*}

Figure~\ref{fig:6} gives the magnetic-moment dynamics for two separate continuous-wave CPL excitations at the respective mode resonances, both starting from the equilibrium structure. The time coordinate is expressed in optical periods, $T_0=1/\nu$. For both modes, the magnetic moment undergoes a pronounced transient overshoot, reaching several times its steady value before its oscillations decay over approximately 3000 optical cycles. This behavior is more pronounced than the predominantly saturating envelope described for benzene.
This comparison shows that the larger molecular framework is accompanied by a different magnetic-moment buildup rather than merely a rescaling of its magnitude. The three systems use the same damping and computational settings, so the difference cannot be assigned to separately chosen damping parameters. The greater mode density of coronene may be relevant to its transient behavior, but identifying the participating modes requires a mode-resolved analysis. The upper insets of Fig.~\ref{fig:6} show the degenerate orthogonal modes within each of the two $E_{1u}$ pairs, and the lower insets show the corresponding closed atomic trajectories. Hydrogen atoms follow particularly flattened ellipses.

Despite its larger number of vibrational degrees of freedom, coronene has a smaller magnetic moment per molecule than benzene. In this model, each atomic contribution depends on the effective charge, vibrational frequency, and oriented area of the trajectory. Hydrogen's larger charge-to-mass ratio favors its driven response. Then under the same mode, driving, and damping, hydrogen will have a larger trajectory area, thereby dominating the total mass-specific magnetic moment of the molecule. From benzene to coronene, the H:C atomic ratio decreases from $1:1$ to $1:2$. Together with coronene's greater molecular mass and smaller total magnetic moment, this implies a smaller magnetic moment per unit mass, suggesting that atomic composition and mode displacements jointly influence the response.

\begin{figure*}[tp]
 \centering
 \includegraphics[width=\JCPwidefigurewidth]{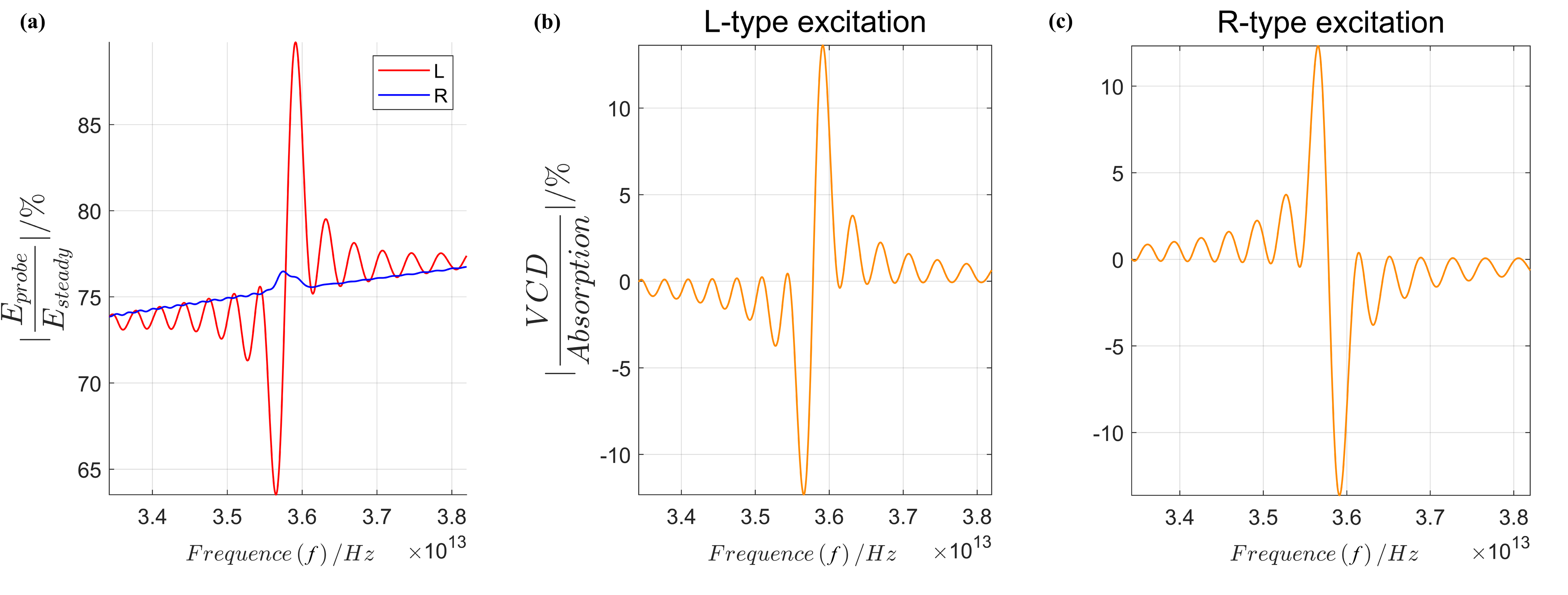}
 \caption{Transient VCD spectra of coronene near mode A at $35.8\,\mathrm{THz}$. (a) Relative spectra for left- and right-handed probes following left-handed pumping. (b) and (c) VCD spectra after resonant left- and right-handed pumping, respectively.}
 \label{fig:7}
\end{figure*}

Figure~\ref{fig:7} presents the transient VCD response near the $35.8\,\mathrm{THz}$ resonance, using the benzene pump--probe procedure. Following a left-handed pump, the same-helicity probe response has pronounced oscillatory structure. The opposite-helicity response remains near the lower baseline except for a small resonance feature. We interpret this difference as preferential coupling to the prepared left-handed vibrational state.
For left-handed pumping, the spectrum in Fig.~\ref{fig:7}(b) has a bisignate oscillatory envelope that changes sign across resonance: negative values below resonance, a zero crossing near resonance, and positive values above it. The oscillatory peaks decrease with detuning. Right-handed pumping preserves the line shape while reversing every sign [Fig.~\ref{fig:7}(c)]. Thus, the negative--positive or positive--negative sign sequence is correlated with the pump helicity.

The observation-window factor in Appendix~\ref{app:vcd} provides a general reference for spectral-shape analysis. Here, $\tau$ is the probe-window duration and $\Gamma$ is the scalar damping rate of the selected mode. Weak damping, $\Gamma/\omega_0\ll1$, is compatible with either a short or a long probe window. For $\Gamma\tau\ll1$, the truncated oscillation produces a sinc-like window factor; for $\Gamma\tau\gg1$, the factor approaches a complex Lorentzian with absorptive and dispersive components. The window factor is not the full mean-energy difference, which also depends on the probe-response amplitude and the relative phase of the residual and driven motion. The three systems use the same damping and pump--probe settings, so these window limits alone do not explain their actual spectral differences. We report the displayed line shapes as calculated results without assigning an unverified molecule-specific origin.

\subsection{Methane}\label{sec:methane}
\begin{figure*}[tp]
 \centering
 \includegraphics[width=\JCPwidefigurewidth]{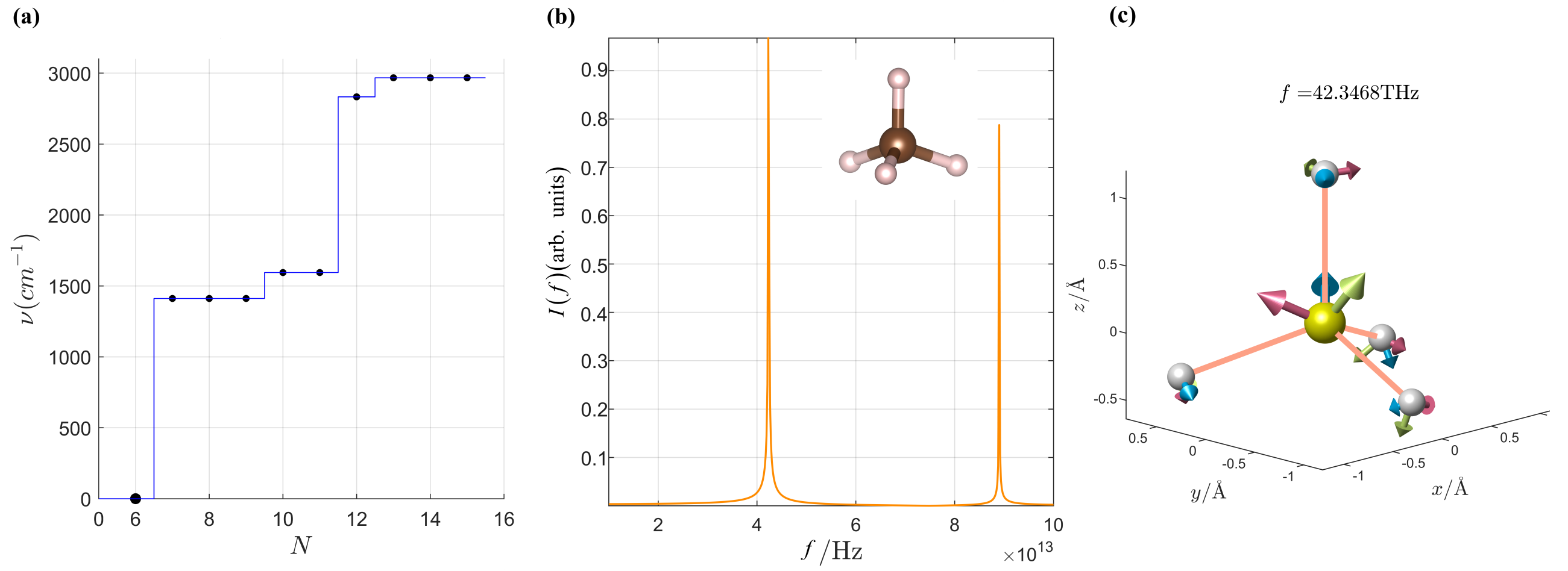}
 \caption{(a) Vibrational levels of methane. (b) Infrared absorption spectrum; inset: molecular structure. (c) Three-dimensional representation of the triply degenerate $T_2$ asymmetric bending mode at $42.3\,\mathrm{THz}$, including its orthogonal components and their superposition into chiral trajectories.}
 \label{fig:8}
\end{figure*}

Methane, $\mathrm{CH}_4$, is a prototype molecule of $T_d$ symmetry, with a central C atom and four equivalent H atoms in a regular tetrahedral arrangement [Fig.~\ref{fig:8}(b)]. Its nonplanar geometry imposes different symmetry constraints from those of benzene and coronene. The five-atom molecule has $3N-6=9$ internal vibrational degrees of freedom, with representation $A_1\oplus E\oplus2T_2$. The calculated modes comprise an $A_1$ symmetric stretch near $87.5\,\mathrm{THz}$, an $E$ bend near $46.0\,\mathrm{THz}$, a $T_2$ asymmetric bend near $42.3\,\mathrm{THz}$, and a $T_2$ asymmetric stretch near $90.5\,\mathrm{THz}$. Only the $T_2$ channels are infrared active. For the high-frequency stretching mode, the atomic displacement components are nearly collinear along the C--H axes; we therefore identify the $42.3\,\mathrm{THz}$ asymmetric bend as the chiral-active mode. This distinction parallels the difference between bending and high-frequency stretching pairs in benzene.
Figure~\ref{fig:8}(a) orders the nine vibrational eigenvalues by frequency. With the six rigid-body modes included in the indexing, modes 7--9 form the $T_2$ bending triplet, modes 10--11 form the infrared-inactive $E$ doublet, mode 12 is the infrared-inactive $A_1$ stretch, and modes 13--15 form the infrared-active $T_2$ stretching triplet. Methane therefore provides a comparatively sparse spectrum for examining triply degenerate dynamics. Its infrared spectrum [Fig.~\ref{fig:8}(b)] contains two peaks, with the $42.3\,\mathrm{THz}$ bending peak substantially stronger than the $90.5\,\mathrm{THz}$ stretching peak. We associate this difference with a larger dipole variation during asymmetric bending.

\begin{figure*}[t]
 \centering
 \includegraphics[width=\JCPwidefigurewidth]{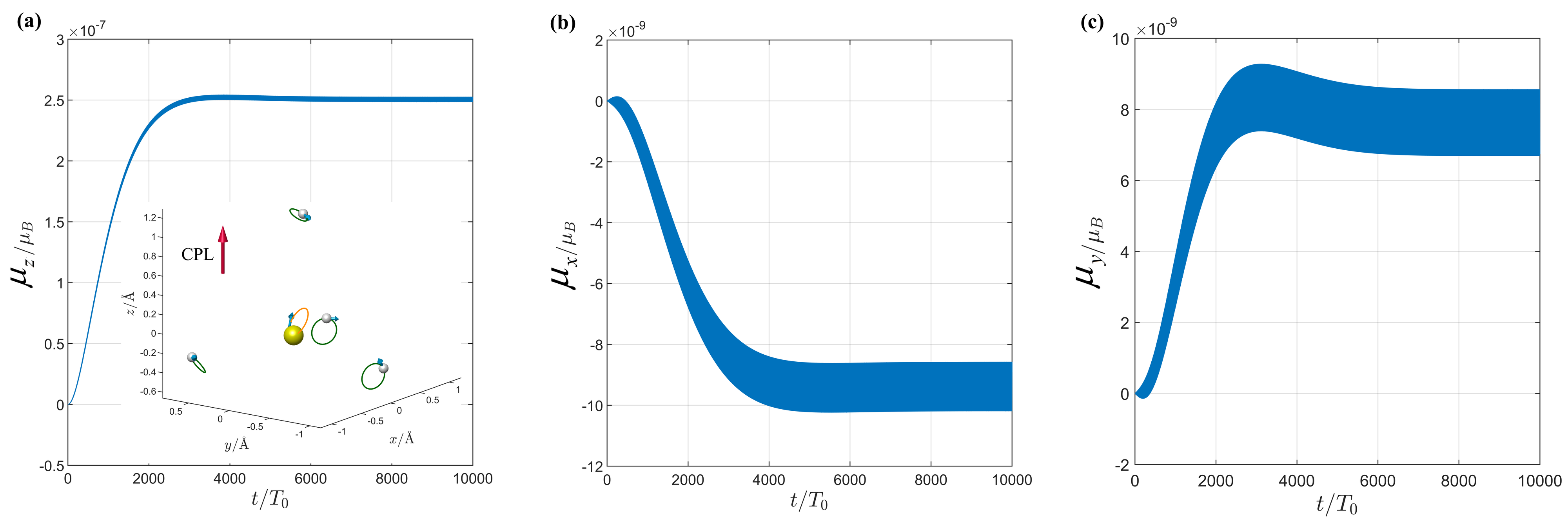}
 \caption{Magnetic-moment evolution of methane under resonant $z$-incident CPL at $42.3\,\mathrm{THz}$. (a) Longitudinal component $\mu_z(t)$, with a saturating envelope and optical-cycle oscillations; inset: excitation geometry and steady atomic trajectories. (b) and (c) Transverse components $\mu_x(t)$ and $\mu_y(t)$, showing oscillations beyond 3000 optical cycles.}
 \label{fig:9}
\end{figure*}

\begin{figure*}[t]
 \centering
 \includegraphics[width=\JCPwidefigurewidth]{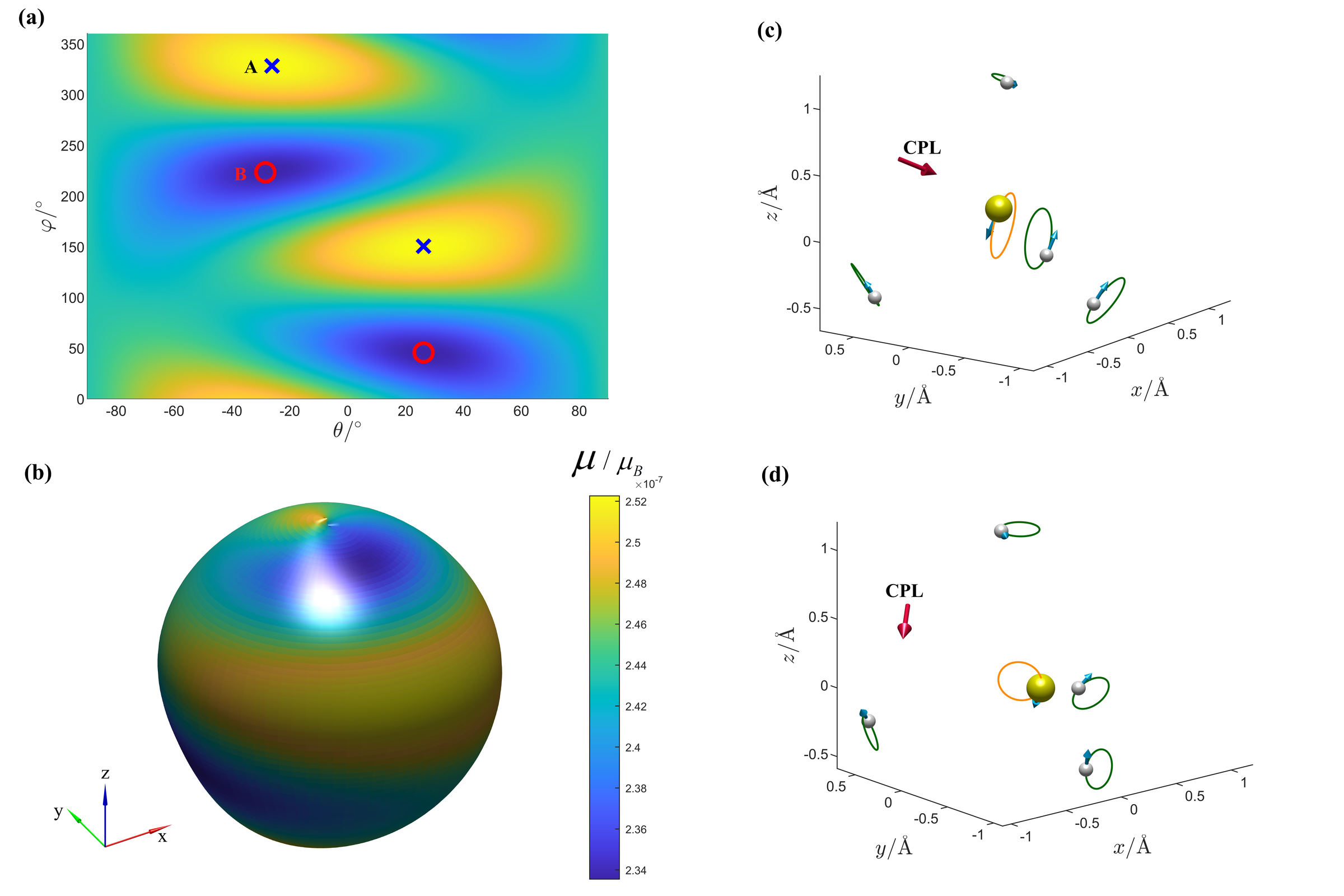}
 \caption{Angular dependence of the magnetic-moment magnitude of methane under CPL excitation at $42.3\,\mathrm{THz}$. (a) Contour map versus the incidence-direction parameters $(\theta,\phi)$; A and B mark the reported maximum and minimum. (b) Three-dimensional polar representation. (c) and (d) Atomic trajectories at A and B, respectively. The calculated response shows angular modulation.}
 \label{fig:10}
\end{figure*}

\begin{figure*}[t]
 \centering
 \includegraphics[width=\JCPwidefigurewidth]{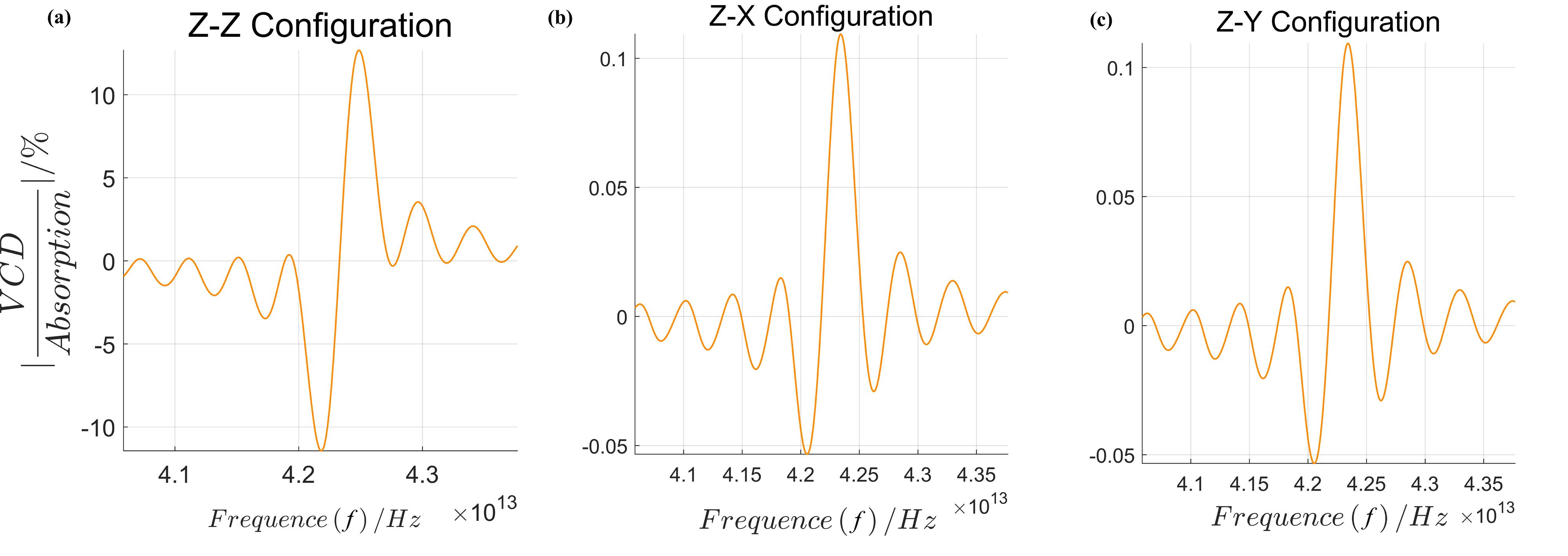}
 \caption{Transient VCD spectra of methane near the $42.3\,\mathrm{THz}$ $T_2$ bending mode after left-handed pumping along $z$. (a) $z$-pump/$z$-probe configuration. (b) $z$-pump/$x$-probe configuration. (c) $z$-pump/$y$-probe configuration.}
 \label{fig:11}
\end{figure*}

The three orthogonal components of the bending triplet transform as $x$, $y$, and $z$ and are denoted $Q_x$, $Q_y$, and $Q_z$. Each is a collective displacement pattern of all atoms, rather than the displacement of an individual atom along one axis. Under a linearly polarized field, an individual component executes linearly polarized motion. CPL propagating along $z$ directly drives the two field-compatible components, $Q_x$ and $Q_y$, with a relative phase of $\pi/2$, producing closed elliptical atomic trajectories.

We further find that the third component, $Q_z$, is indirectly excited through coupling to $Q_x$ and $Q_y$. Although it is orthogonal to the $z$-incident field and is not directly driven, the coupling transfers coherent energy into it and establishes a finite amplitude and phase. In that interpretation, the superposition of three components at a degenerate frequency produces closed ellipses in tilted planes.The four H atoms move around their equilibrium positions, while the C atom follows a smaller, oppositely phased trajectory so that the center of mass is conserved. Circulating positive H and negative C charges yield calculated longitudinal and transverse magnetic-moment components, $\mu_z$ and $\mu_x,\mu_y$. Unlike the two separately resonant chiral pairs in benzene, methane's $Q_x$, $Q_y$, and $Q_z$ belong to the same triply degenerate $T_2$ manifold at $42.3\,\mathrm{THz}$. At a common driving frequency, the local displacements of the components that are actually excited determine the trajectory geometry; this does not establish coherent energy exchange among all three components.

\begingroup\clubpenalty=10000
Figure~\ref{fig:9} shows all three Cartesian components of the photoinduced magnetic moment under continuous resonant CPL propagating along $z$. The molecule starts at its equilibrium geometry. The longitudinal component rises with a predominantly monotonic envelope and approaches a positive steady value after approximately 3000 optical cycles. At $42.3\,\mathrm{THz}$, $T_0=1/\nu\approx23.6\,\mathrm{fs}$, so this interval is approximately $70\,\mathrm{ps}$. Many cycles of high-frequency oscillation are superimposed on the slowly varying envelope, with residual oscillations visible at late times. The envelope does not exhibit the pronounced overshoot of coronene.
\par\endgroup

Figures~\ref{fig:9}(b) and \ref{fig:9}(c) show sustained temporal oscillations of the transverse magnetic-moment components ($\mu_x$ and $\mu_y$). This coupled dynamics is directly manifested as large, asynchronous, underdamped oscillations of the three magnetic moment components. By comparison, the selected modes of the planar molecules considered above do not produce an out-of-plane magnetic response.

Figure~\ref{fig:10} examines the dependence of the steady total magnetic moment on incidence direction. The angles $\theta$ and $\phi$ parameterize the incidence directions shown in the figure. The contour plot identifies a maximum at point A and a minimum at point B, showing angular modulation of the response. The angular pattern should be analyzed in relation to tetrahedral symmetry and the direction-dependent excitation weights of the three $T_2$ components, but molecular geometry alone does not determine the modulation quantitatively.
In Fig.~\ref{fig:10}(b), radial distance and color encode the $\lvert\vect\mu\rvert$ magnitude. The departures from a sphere visualize the angular anisotropy of the response. Changing the molecular orientation, or equivalently the incidence direction, therefore changes the magnitude and direction of the calculated photoinduced magnetic moment at fixed optical frequency and intensity. The trajectories at A are described as more nearly circular and enclosing larger areas, whereas
those at B are flatter and enclose smaller areas [Figs.~\ref{fig:10}(c) and \ref{fig:10}(d)]. We attribute the approximately $10\%$ modulation to this orbital difference. We find no incidence direction with a vanishing moment, in contrast to the grazing-incidence limit of benzene, and associate this behavior with the availability of multiple $T_2$ components.

\begin{figure*}[t]
 \centering
 \includegraphics[width=\JCPwidefigurewidth]{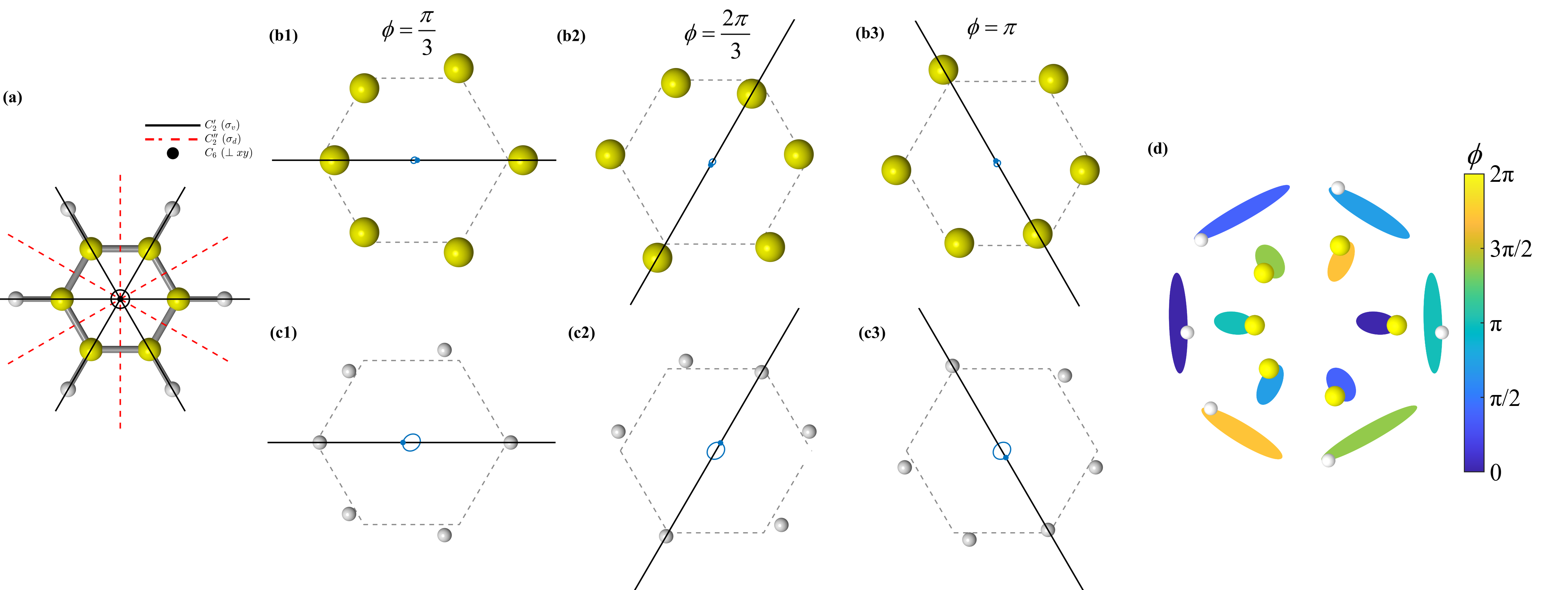}
 \caption{Symmetry breaking and phase-dependent symmetry elements in the driven benzene state. (a) Equilibrium structure and $D_{6h}$ symmetry elements. (b1)--(b3) C-atom configurations at the three indicated optical phases, $\pi/3$, $2\pi/3$, and $\pi$, with the surviving vertical mirror indicated. (c1)--(c3) Corresponding H-atom configurations. A scale factor of 1.79 is used to compare H and C rings with radii $2.49$ and $1.39\,\text{\AA}$, respectively. (d) Full-period trajectories of all atoms, colored by their initial phase $\phi$.}
 \label{fig:12}
\end{figure*}

Figure~\ref{fig:11} presents the transient VCD spectra after preparation by left-handed CPL propagating along $z$. After the pump is removed, left- and right-handed probes are applied along $z$, $x$, or $y$. The $z$-probe spectrum [Fig.~\ref{fig:11}(a)] has a nearly antisymmetric bisignate profile, negative below $42.3\,\mathrm{THz}$ and positive above it, with a zero crossing near resonance. This shape qualitatively resembles the calculated coronene spectrum. Subject to the limitations discussed above, the general window and phase analysis in Appendix~\ref{app:vcd} does not yet establish why these two actual molecules display this shape, and a bisignate appearance alone does not determine $\Gamma\tau$.

The transverse-probe spectra [Figs.~\ref{fig:11}(b) and \ref{fig:11}(c)] have approximately sinc-like envelopes. But the whole lineshape is asymmetric between the left and right lobes. An $x$-directed probe can couple to $Q_y$ and $Q_z$, while a $y$-directed probe can couple to $Q_x$ and $Q_z$.The probe actually detect the 3D vibration mode in one projection plane.  Similar to Appendix A, the frequency-dependent phase relationships between the residual motion and the probe-driven response also affect the differential signal. Although the mode couplings in three directions generate the VCD line shapes, one single VCD signal can not reproduce all these couplings.

\subsection{Group analysis: From equilibrium point groups to dynamical symmetry}\label{sec:symmetry}
The previous results suggest two distinct requirements within the resonant normal-mode picture. One is the symmetry-protected degeneracy, and the other is the noncollinear atomic displacement component with an appropriate relative phase. The first condition follows from a multidimensional irreducible representation; the second condition must be examined from the atomic eigenvectors within each candidate manifold. Exactly because of the independence of these two conditions, the modes belonging to $E_1u$ can be either chiral or only linearly polarized. This also leads to the different situations of degenerate modes in the previous three molecules. Appendix~\ref{app:group} gives the equilibrium selection rules.

An equilibrium point group alone does not characterize a driven and periodic state; one must also examine how the atomic configuration evolves with time. Figure~\ref{fig:12} illustrates this analysis for benzene: equilibrium symmetry elements, representative instant C and H configurations at the indicated optical phases $\pi/3$, $2\pi/3$, and $\pi$, and the complete steady trajectories colored by initial phase.

At equilibrium, Fig.~\ref{fig:12}(a) includes the three $C_2'$ axes, three $C_2''$ axes, and the perpendicular sixfold rotation axis. The driven displacement changes the instantaneous configuration and its residual symmetry. In each illustrated planar configuration, the marked mirror line passes through a pair of opposite carbon atoms and the molecular center. Reflection across this line exchanges symmetry-related atoms within the molecular plane. The illustrated H-atom configurations share the same mirror line. The C and H rings are rescaled for visual comparison, and the indicated mirror direction rotates cynchronously with the CPL phases.

Figure~\ref{fig:12}(d) shows the complete trajectories of all 12 atoms over one period. Opposite C atoms differ in phase by $\pi$, and each H atom has the same phase as the opposite C atom. Under continuous circularly polarized driving, the atoms move along periodic trajectories. The instantaneous configuration and its symmetry elements change with the optical phase and repeat after one period. The instantaneous symmetry elements therefore form a time-dependent pattern.

The periodic pattern in Fig.~\ref{fig:12}(d) motivates a description in terms of combined spatial and temporal operations. For continuous CPL with period $T=2\pi/\omega$, these operations pair a rotation by $\delta$ about the pump axis with a time advance of $\delta/\omega$. The resulting continuous family $\{R(\delta)\mathcal{T}(\delta/\omega)\}$ leaves the entire driven system invariant. This symmetry therefore connects the instantaneous symmetry breaking described above with recovery over a full period. It also distinguishes the dynamical symmetries of the two driving polarizations: linearly polarized driving retains only a fixed spatial operation, whereas CPL supports a continuous dynamical rotation group. The alignment of the steady magnetic moment and the VCD response with the pump axis [Figs.~\ref{fig:2}, \ref{fig:4}, and \ref{fig:9}(a)] is a direct manifestation of this symmetry. These observables connect the dynamical symmetry of each periodic state to the handedness prepared and detected in the writing-readout scheme.

Vibrational enantiomers acquire a more precise definition within this framework. At equilibrium, all mirror planes belong to the molecular point group, whose spatial operations relate left- and right-handed CPL by reflection. The two handednesses are therefore equivalent and indistinguishable at equilibrium. Time reversal, denoted by TI, distinguishes the driven left- and right-handed steady states. The two oppositely rotating vibrational enantiomers are related by TI, rather than by any spatial operation of the equilibrium point group acting alone. This gives an one-to-one correspondence among pump helicity, the rotation direction of the vibrational steady state, and the sign of transient VCD [Figs.~\ref{fig:4}, \ref{fig:7}, and \ref{fig:11}], allowing readout without additional calibration.

These links between molecular symmetry and handedness suggest a systematic classification of driven chiral states. Such classification would extend the static description to dynamic and nonequilibrium regimes.

\section{Conclusions}\label{sec:conclusions}
We have developed a molecular description of the vibrational IFE and its transient VCD readout using DFTB+-derived force constants and damped normal-mode dynamics. CPL can prepare rotating nuclear motion in suitable degenerate infrared-active modes of otherwise achiral molecules. The circulation of atomic charges produces an effective magnetic moment, while a subsequent helicity-dependent probe provides an optical measure of the prepared vibrational state.

The calculations for benzene, coronene, and methane exhibit different temporal and spectral responses. The planar $D_{6h}$ molecules provide doubly degenerate $E_{1u}$ channels, while methane provides a triply degenerate $T_2$ manifold. The reported behavior includes a predominantly saturating benzene response, transient overshoot in coronene, and longitudinal and transverse magnetic-moment oscillations in methane. Reversing the pump helicity reverses the sign of the transient dichroic signal. Equilibrium selection rules, degeneracy, and local displacement geometry provide complementary criteria for identifying candidate vibrational channels. The differences among these observations do not, by themselves, establish the specific modal mechanism in each system.

The proposed writing-readout scheme connects molecular symmetry, nuclear motion, magnetic response, and transient optical chirality. It provides a molecular counterpart to phonon-driven magnetism in crystals and suggests that excitation geometry and chemical composition can control the response. The present treatment is based on damped, near-resonant normal modes. A fuller description is needed for explicit intermode coupling, anharmonicity, and strong-field electron--vibration dynamics. Extensions to larger fused aromatic molecules, $O_h$ or $I_h$ systems with three-dimensional representations, and intrinsically chiral molecules may clarify how structural and vibrational chirality cooperate or compete.

\begin{acknowledgments}
This work was supported by the National Natural Science Foundation of China (Grant No. 12547101). Partial support from the Fundamental Research Fund for the Central Universities (Grant No. 2025CDJ-IAISYB-035) and the New Chongqing Youth Innovative Talent Project (Grant No. CSTB2025YITP-QCRCX0069) is also acknowledged.
\end{acknowledgments}

\section*{Author declarations}
\subsection*{Conflict of interest}
The authors have no conflicts to disclose.

\subsection*{Author contributions}
\textbf{Hao-Wen Xu:} Conceptualization; Investigation; Formal analysis; Writing -- original draft; Writing -- review \& editing.
\textbf{Hang Xie:} Conceptualization; Investigation; Formal analysis; Writing -- original draft; Writing -- review \& editing.
\textbf{Xiao-Long Yang:} Investigation; Formal analysis (damping-coefficient calculations and analysis).
\textbf{Cheng Xu:} Investigation; Formal analysis (damping-coefficient calculations and analysis).
\textbf{Zi-Yang Xu:} Investigation; Formal analysis (angular momentum and magnetic moment analysis).

\section*{Data availability}
The data that support the findings of this study are available from the corresponding author upon reasonable request.

\appendix
\begingroup

\section{Analytical model of transient VCD}\label{app:vcd}

This appendix develops an analytical model of doubly degenerate vibrations to describe the transient VCD readout of a pump-prepared state. The model considers a single degenerate frequency; contributions from modes at different frequencies are treated numerically.

\subsection{Model and potential energy}\label{app:vcd:1}

Following the mass-weighted mode decomposition in the main text, the external driving force is first expressed as a mass-weighted vector $\widetilde{\vect F}$ and projected onto the normal coordinates as $\vect G=\mathbf U^{\mathrm T}\widetilde{\vect F}$. We select a doubly degenerate, infrared-active frequency subspace and denote its two mass-weighted normal coordinates by $Q_1$ and $Q_2$. The equations of motion in this subspace are
\begin{equation}\label{eq:A1}
\ddot Q_i+\Gamma\dot Q_i+\omega_0^2Q_i=G_i(t),
\qquad i=1,2.
\end{equation}

Degeneracy gives the two components the same restoring coefficient, and their potential energy is
\begin{equation}\label{eq:A2}
U=\frac12 kQ_1^2+\frac12 kQ_2^2,
\qquad k=\omega_0^2.
\end{equation}
The value of $k$ may differ between frequency subspaces but remains the same within the degenerate subspace considered here. CPL has driving components of equal amplitude along two orthogonal directions. In this two-dimensional isotropic reduced model, we further assume equal amplitudes of the effective modal driving forces and equal damping in the two directions. Let $\hat{\vect e}_1$ and $\hat{\vect e}_2$ denote the two orthogonal electric-field directions; for the in-plane $E_{1u}$ degenerate pair, these are chosen along $x$ and $y$, respectively. With $\vect p$ denoting the electric dipole moment, we define
\begin{equation}\label{eq:A3}
\begin{aligned}
G_i(t)&=g\,\hat{\vect e}_i\cdot\vect E(t),\qquad i=1,2,\\
g&=\left|\frac{\partial\vect p}{\partial Q}\right|.
\end{aligned}
\end{equation}

The CPL electric field propagating along $z$ with helicity $s=\pm1$ is
\begin{equation}\label{eq:A4}
\vect E(t)=E_0[\hat{\vect x}\cos\omega t+s\hat{\vect y}\sin\omega t].
\end{equation}
The two effective driving-force components are therefore
\begin{equation}\label{eq:A5}
G_1(t)=gE_0\cos\omega t,
\qquad
G_2(t)=sgE_0\sin\omega t.
\end{equation}

\subsection{Steady-state motion under pumping}\label{app:vcd:2}

For harmonic driving at angular frequency $\omega$, we write
\begin{equation}\label{eq:A6}
\begin{aligned}
Q_i(t)&=\Rea[Z_i e^{i\omega t}],\\
G_1(t)&=G_0\Rea[e^{i\omega t}],\\
G_2(t)&=sG_0\Rea[-i e^{i\omega t}],
\qquad G_0=gE_0.
\end{aligned}
\end{equation}
Substituting Eq.~\eqref{eq:A6} into Eq.~\eqref{eq:A1}, we define the response denominator
\begin{equation}\label{eq:A7}
D(\omega)=\omega_0^2-\omega^2+i\Gamma\omega.
\end{equation}
The two complex amplitudes are then
\begin{equation}\label{eq:A8}
Z_1(\omega)=\frac{G_0}{D(\omega)},
\qquad
Z_2(\omega)=-is\frac{G_0}{D(\omega)}.
\end{equation}
The complex amplitudes have equal magnitudes and a relative phase of $\pm\pi/2$, determined by $s$. We define the response amplitude
\begin{equation}\label{eq:A9}
A(\omega)=\frac{G_0}{|D(\omega)|}
=\frac{G_0}{\sqrt{(k-\omega^2)^2+\Gamma^2\omega^2}},
\end{equation}
and the response phase lag
\begin{equation}\label{eq:A10}
\theta(\omega)=
\begin{cases}
\arctan\!\left(\dfrac{\Gamma\omega}{k-\omega^2}\right),&0<\omega<\omega_0,\\[5pt]
\pi/2,&\omega=\omega_0,\\[3pt]
\pi+\arctan\!\left(\dfrac{\Gamma\omega}{k-\omega^2}\right),&\omega>\omega_0.
\end{cases}
\end{equation}
Here, $\arctan$ takes its principal value and $\Gamma>0$; this definition ensures that $\theta$ is continuous within $0<\theta<\pi$. The time-domain steady-state solution is
\begin{equation}\label{eq:A11}
Q_1=A\cos(\omega t-\theta),
\qquad
Q_2=sA\sin(\omega t-\theta).
\end{equation}
The trajectory in the normal-coordinate plane is a circle of radius $A$, whose direction of rotation reverses with the pump helicity. After reconstruction using the mode eigenvectors, the Cartesian trajectory of each atom is generally elliptical.

\subsection{Squared displacement and vibrational chirality}\label{app:vcd:3}

Averaging the steady-state motion over one period gives
\begin{equation}\label{eq:A12}
\langle Q_1^2\rangle=\langle Q_2^2\rangle=\frac12 A^2.
\end{equation}
The average potential energy is therefore
\begin{equation}\label{eq:A13}
\langle U\rangle=\frac{k}{2}
\bigl(\langle Q_1^2\rangle+\langle Q_2^2\rangle\bigr)
=\frac12 kA^2.
\end{equation}
This energy is independent of $s$, whereas the phase correlation between the two coordinates gives
\begin{equation}\label{eq:A14}
\langle Q_1\dot Q_2-Q_2\dot Q_1\rangle=sA^2\omega.
\end{equation}
This vibrational angular momentum changes sign when the pump helicity is reversed. During the pump--probe process, the associated phase information enters the energy difference through interference between the residual motion and the probe response, enabling a chirality readout.

\subsection{Decomposition of motion during probing}\label{app:vcd:4}

At $t=0$, the pump is switched off and a weak CPL probe is applied. The pump frequency is denoted by $\omega_{\mathrm p}$ and the scanned probe frequency by $\omega$. During probing, the pump-prepared state is characterized by a fixed amplitude $A$, whereas the driven amplitude $a(\omega)$ varies with the probe frequency. Under weak damping, the total motion is decomposed as
\begin{equation}\label{eq:A15}
Q_i(t)=Q_i^{\mathrm f}(t)+Q_i^{\mathrm d}(t).
\end{equation}
For resonant pumping and equal initial phases of the external fields, we take the switching time as the time origin. Retaining the $\pi/2$ phase lag of the resonant response, the residual motion is written as
\begin{equation}\label{eq:A16}
\begin{aligned}
Q_1^{\mathrm f}(t)&\approx Ae^{-\Gamma t/2}\cos(\omega_0t-\pi/2),\\
Q_2^{\mathrm f}(t)&\approx sAe^{-\Gamma t/2}\sin(\omega_0t-\pi/2).
\end{aligned}
\end{equation}
The steady-state response driven by the probe is
\begin{equation}\label{eq:A17}
\begin{aligned}
Q_1^{\mathrm d}(t)&=a(\omega)\cos[\omega t-\theta(\omega)],\\
Q_2^{\mathrm d}(t)&=s'a(\omega)\sin[\omega t-\theta(\omega)],\\
a(\omega)&=\frac{G_{\mathrm{pr}}}{|D(\omega)|},
\qquad G_{\mathrm{pr}}=gE'_0.
\end{aligned}
\end{equation}
Here, $s'=\pm1$ is the probe helicity. These expressions correspond to a zero reference phase difference between the two states at resonance; Sec.~\ref{app:vcd:5} generalizes them to an arbitrary phase difference. We set the field-amplitude ratio to $E'_0/E_0=0.01$ and treat the pump-prepared state as a fixed background. The transient at probe onset is neglected for now; its correction and comparison are presented in Sec.~\ref{app:vcd:8}.

One probe period is $T=2\pi/\omega$, and the total observation time is $\tau=100T$. The time average is defined as $\langle f\rangle=\tau^{-1}\int_0^\tau f(t)\,dt$. Writing $(Q^{\mathrm f})^2=\sum_i(Q_i^{\mathrm f})^2$ and $(Q^{\mathrm d})^2=\sum_i(Q_i^{\mathrm d})^2$, we obtain
\begin{equation}\label{eq:A18}
\begin{aligned}
\langle Q_1^2+Q_2^2\rangle={}&\langle(Q^{\mathrm f})^2\rangle
+\langle(Q^{\mathrm d})^2\rangle\\
&+2\langle Q_1^{\mathrm f}Q_1^{\mathrm d}
+Q_2^{\mathrm f}Q_2^{\mathrm d}\rangle.
\end{aligned}
\end{equation}
The first two terms are $A^2\langle e^{-\Gamma t}\rangle$ and $a^2$, respectively. Near resonance, we average over the fast oscillations in the cross terms, retaining the difference-frequency contribution and neglecting the sum-frequency contribution. Defining
\begin{equation}\label{eq:A19}
W(\Delta)=\frac1\tau\int_0^\tau e^{-\Gamma t/2}e^{i\Delta t}\,dt,
\qquad \Delta=\omega-\omega_0,
\end{equation}
we obtain
\begin{equation}\label{eq:A20}
\begin{aligned}
\langle Q_1^{\mathrm f}Q_1^{\mathrm d}\rangle
&\approx\frac{Aa}{2}\Rea[W(\Delta)e^{-i(\theta-\pi/2)}],\\
\langle Q_2^{\mathrm f}Q_2^{\mathrm d}\rangle
&\approx ss'\frac{Aa}{2}\Rea[W(\Delta)e^{-i(\theta-\pi/2)}].
\end{aligned}
\end{equation}
Writing $x^2=Q_1^2+Q_2^2$, the mean squared displacement is
\begin{equation}\label{eq:A21}
\begin{aligned}
\langle x^2\rangle\approx{}&\langle(Q^{\mathrm f})^2\rangle
+\langle(Q^{\mathrm d})^2\rangle\\
&+Aa\Rea[W(\Delta)e^{-i(\theta-\pi/2)}]\\
&+ss'Aa\Rea[W(\Delta)e^{-i(\theta-\pi/2)}].
\end{aligned}
\end{equation}
Only the final term depends on $s'$ and thus determines the difference in average elastic energy between opposite probe helicities.

\subsection{Relative phase of the two states and transient VCD}\label{app:vcd:5}

Let the initial phase of the residual vibrational state be $\psi_{\mathrm f}$ and that of the probe-driven vibrational state be $\psi_{\mathrm d}(\omega)=\varphi_{\mathrm{pr}}-\theta(\omega)$, where $\varphi_{\mathrm{pr}}$ is the initial phase of the probe field. We define
\begin{equation}\label{eq:A22}
\phi\equiv\psi_{\mathrm f}-\psi_{\mathrm d}(\omega_0)
=\psi_{\mathrm f}-\varphi_{\mathrm{pr}}+\frac{\pi}{2}.
\end{equation}
\(\phi\) is the phase difference between the two vibrational states at the switching time under resonant conditions. This phase difference serves as the reference phase in all subsequent calculations. Possible origins include the pump-response lag, free evolution after the pump is switched off, and polarization-induced or optical-path phase shifts experienced by the two excitations. For steady-state pumping, $\psi_{\mathrm f}\approx\varphi_{\mathrm p}-\theta(\omega_{\mathrm p})$; pump detuning and excitation timing can therefore change the relative phase of the two states.

Away from resonance, the initial phase difference between the two states is $\phi+\theta(\omega)-\pi/2$. Substituting this phase difference into Eqs.~\eqref{eq:A20} and~\eqref{eq:A21}, the helicity-independent squared terms cancel upon taking the difference, giving
\begin{equation}\label{eq:A23}
\begin{aligned}
S(\omega,\phi)&\equiv\VCD\\
&=\frac{k}{2}\left[\langle x^2\rangle_{s'=+1}
-\langle x^2\rangle_{s'=-1}\right]\\
&\approx ksAa(\omega)\Rea\!\left[
W(\Delta)e^{-i[\phi+\theta(\omega)-\pi/2]}\right].
\end{aligned}
\end{equation}
This signal is proportional to the product of the residual pump-induced amplitude and the probe amplitude, and changes sign when the pump helicity $s$ is reversed.

\subsection{Even and odd line-shape components for a finite time window}\label{app:vcd:6}

Letting $\alpha=\Gamma/2$ and $r=e^{-\alpha\tau}$, integration of Eq.~\eqref{eq:A19} gives
\begin{equation}\label{eq:A24}
W(\Delta)=\frac{1-r e^{i\Delta\tau}}{\tau(\alpha-i\Delta)}.
\end{equation}
Using $a e^{-i\theta}=G_{\mathrm{pr}}/D$, Eq.~\eqref{eq:A23} can be written as
\begin{equation}\label{eq:A25}
S(\omega,\phi)\approx ksAG_{\mathrm{pr}}
\Rea\!\left[\frac{i e^{-i\phi}W(\Delta)}{D(\omega)}\right].
\end{equation}
For $\alpha,|\Delta|\ll\omega_0$, the response denominator expands as
\begin{equation}\label{eq:A26}
\begin{aligned}
D(\omega)&=2\omega_0(-\Delta+i\alpha)-\Delta^2+i2\alpha\Delta\\
&\approx2\omega_0(-\Delta+i\alpha).
\end{aligned}
\end{equation}
Substituting this expression and Eq.~\eqref{eq:A24} into Eq.~\eqref{eq:A25}, while retaining the full oscillatory dependence on $\Delta\tau$, gives
\begin{equation}\label{eq:A27}
\begin{aligned}
S(\omega,\phi)\approx{}&
\frac{ksAG_{\mathrm{pr}}}{2\omega_0\tau(\Delta^2+\alpha^2)}\\
&\times\bigl\{[1-r\cos(\Delta\tau)]\cos\phi\\
&\hspace{2em}{}-r\sin(\Delta\tau)\sin\phi\bigr\}.
\end{aligned}
\end{equation}
We define the dimensionless detuning $z=\Delta\tau$, the decay parameter $u=\alpha\tau$, and
\begin{equation}\label{eq:A28}
\begin{aligned}
f_{\mathrm e}^{(0)}(z,u)&=\frac{1-e^{-u}\cos z}{z^2+u^2},\\
f_{\mathrm o}^{(0)}(z,u)&=-\frac{e^{-u}\sin z}{z^2+u^2},
\end{aligned}
\end{equation}
so that the signal becomes
\begin{equation}\label{eq:A29}
\begin{aligned}
S\approx{}&\frac{ksAG_{\mathrm{pr}}\tau}{2\omega_0}
\bigl[f_{\mathrm e}^{(0)}(z,u)\cos\phi\\
&\hspace{5.5em}{}+f_{\mathrm o}^{(0)}(z,u)\sin\phi\bigr].
\end{aligned}
\end{equation}
For fixed $u$, $\tau$ and $\phi$, $f_{\mathrm e}^{(0)}$ and $f_{\mathrm o}^{(0)}$ are even and odd in detuning, respectively. The phase $\phi=0$ selects the even component, whereas $\phi=\pi/2$ selects the odd component; other phases correspond to their superposition. A frequency-dependent phase and observation window can be incorporated into Eq.~\eqref{eq:A27} through $\phi(\omega)$ and $\tau(\omega)$, respectively.

\subsection{Control of the line shape by the relative phase}\label{app:vcd:7}

Figure~\ref{fig:appendix-phase} shows four phase slices for $s=+1$ and a resonant decay parameter $u_0=\Gamma\tau_0/2=1/30$, where $\tau_0=200\pi/\omega_0$, over the detuning range $-0.2\le z\le0.2$. The figure compares a fixed window $\tau=\tau_0$ with a window spanning 100 probe periods, $\tau(\omega)=200\pi/\omega$. For the latter, $\tau(\omega)/\tau_0=1-z/(200\pi)$; accordingly, we use $u(\omega)=u_0[1-z/(200\pi)]$ and retain the window dependence of the prefactor in Eq.~\eqref{eq:A29}. All curves are normalized by the signal at resonance for $\phi=0$, namely $S_{\mathrm{ref}}=S(\omega_0,0)$.

\begin{figure*}[t]
\centering
\includegraphics[width=\JCPwidefigurewidth]{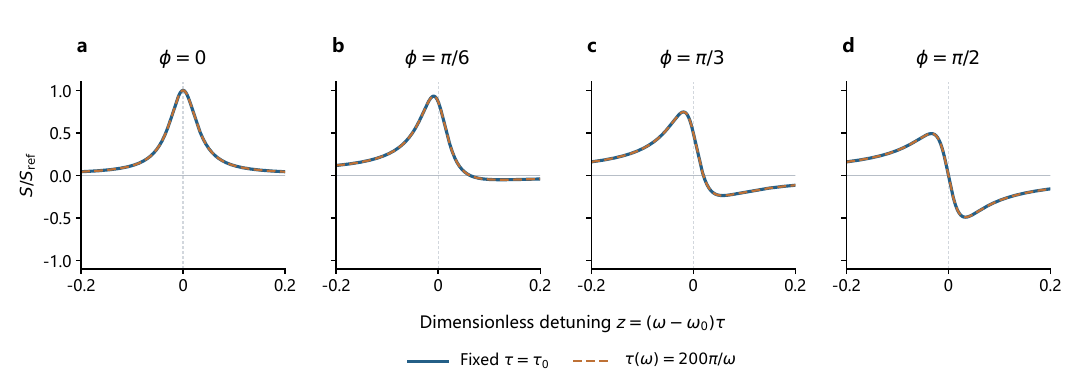}
\caption{Phase-dependent VCD line shapes near resonance. From left to right, $\phi=0,\pi/6,\pi/3,\pi/2$; $u_0=1/30$ and $s=+1$. Solid blue curves represent the fixed window, and dashed orange curves represent the window spanning 100 probe periods. All curves use a common normalization.}
\label{fig:appendix-phase}
\end{figure*}

For the fixed window, the spectrum at $\phi=0$ is symmetric about resonance. As $\phi$ increases, the weight of the odd component grows and the spectrum becomes increasingly asymmetric. At $\phi=\pi/2$, the even component vanishes, and the signal crosses zero at resonance and has opposite signs on either side. The normalized values at the center for the four phases are $1,\sqrt3/2,1/2,0$, respectively. Within the narrow frequency range shown, the curves for the two windows nearly coincide, indicating that the frequency dependence of the window causes only small line-shape deviations. Reversing the pump helicity changes the sign of the entire signal, whereas the phase determines the relative weights of the even and odd components.

\subsection{Continuity correction at probe onset}\label{app:vcd:8}

To describe probe onset, we supplement the steady-state particular solution for the probe with a homogeneous correction that ensures continuity of the initial conditions. At the same order of the weak-damping, near-resonance and difference-frequency averaging approximations, the window factor changes from $W$ to $W-H$, where
\begin{equation}\label{eq:A30}
H=\frac1\tau\int_0^\tau e^{-2\alpha t}\,dt
=\frac{1-e^{-2u}}{2u}.
\end{equation}
The corresponding two line-shape functions are
\begin{equation}\label{eq:A31}
\begin{aligned}
f_{\mathrm e}^{(1)}(z,u)
&=\frac{(1+e^{-2u})/2-e^{-u}\cos z}{z^2+u^2},\\
f_{\mathrm o}^{(1)}(z,u)
&=\frac{z(1-e^{-2u})/(2u)-e^{-u}\sin z}{z^2+u^2}.
\end{aligned}
\end{equation}
Their phase weights remain $\cos\phi$ and $\sin\phi$. Comparison with Eq.~\eqref{eq:A28} gives the respective corrections
\begin{equation}\label{eq:A32}
\begin{aligned}
f_{\mathrm e}^{(1)}-f_{\mathrm e}^{(0)}
&=-\frac{1-e^{-2u}}{2(z^2+u^2)},\\
f_{\mathrm o}^{(1)}-f_{\mathrm o}^{(0)}
&=\frac{z(1-e^{-2u})}{2u(z^2+u^2)}.
\end{aligned}
\end{equation}
For a fixed window, these corrections preserve even and odd parity, respectively, leaving the phase-controlled classification into even, odd and mixed line shapes unchanged. They adjust the peak heights, spectral wings and relative weights of the oscillatory structure. Under the short-window condition $u\ll1\ll\omega_0\tau$, the corrected line-shape functions approach
\begin{equation}\label{eq:A33}
\begin{aligned}
f_{\mathrm e}^{(1)}(z,0)&=\frac{1-\cos z}{z^2}
=\frac12\operatorname{sinc}^2(z/2),\\
f_{\mathrm o}^{(1)}(z,0)&=\frac{z-\sin z}{z^2}.
\end{aligned}
\end{equation}
Here, $\operatorname{sinc}v=\sin v/v$, and the continuous values at $z=0$ are $1/2$ and $0$, respectively. This short-window result gives the limiting line shapes after the continuity correction; selection of the even and odd components by the relative phase remains unchanged.
\endgroup

\section{Group-theoretical classification of molecular vibrations}\label{app:group}
The $3N$ Cartesian displacements of an $N$-atom molecule form a representation of the molecular point group $G$. For an operation $g$, its character equals the number $n(g)$ of atoms left fixed multiplied by the trace of the Cartesian transformation. Writing $\mathbf R_g$ for its action on Cartesian vectors gives
\begin{equation}\label{eq:B1}
 \chi^{3N}(g)=n(g)\operatorname{tr}\mathbf R_g,
 \quad\operatorname{tr}\mathbf R_g=
 \begin{cases}
 1+2\cos\theta_g,&\text{proper},\\
 -1+2\cos\theta_g,&S(\theta_g).
 \end{cases}
\end{equation}
Here $\mathbf R_g$ is the Cartesian transformation matrix. An improper operation $S(\theta_g)$ means a rotation by $\theta_g$ followed by reflection in the plane perpendicular to its axis. A mirror is the $\theta_g=0$ case, with trace 1; inversion is the $\theta_g=\pi$ case, with trace $-3$.
Removing overall translations and rotations gives the vibrational representation. For $D_{6h}$,
$\Gamma_{\mathrm{trans}}=E_{1u}\oplus A_{2u}$ and
$\Gamma_{\mathrm{rot}}=E_{1g}\oplus A_{2g}$; for $T_d$,
$\Gamma_{\mathrm{trans}}=T_2$ and $\Gamma_{\mathrm{rot}}=T_1$.
The multiplicity of an irreducible representation $\mu$ follows from character orthogonality:
\begin{equation}\label{eq:B2}
 n_\mu=\frac1h\sum_j h_j\chi^{\mathrm{vib}}(C_j)[\chi^\mu(C_j)]^*,
\end{equation}
where $h$ is the group order, $h_j$ the size of class $C_j$, and the asterisk denotes complex conjugation. The characters are real for the point groups considered here, so the conjugation leaves the present counts unchanged.

Infrared activity requires a nonzero transition-dipole matrix element $\langle f|\hat{\vect r}|i\rangle$. For the ground-state vibrational transitions considered, the product of the mode and dipole representations must contain the totally symmetric representation:
\begin{equation}\label{eq:B3}
 \Gamma_\mu\otimes\Gamma_{\mathrm{dip}}\supset
 \begin{cases}A_{1g},&D_{6h},\\ A_1,&T_d,
 \end{cases}
 \quad\Longleftrightarrow\quad\text{IR active}.
\end{equation}
The dipole transforms as $A_{2u}(z)$ and $E_{1u}(x,y)$ in $D_{6h}$ and as $T_2$ in $T_d$. Raman-active representations are contained in the quadratic Cartesian functions; for $D_{6h}$ these include $A_{1g}$, $E_{1g}$, and $E_{2g}$. These rules identify the infrared channels in Figs.~\ref{fig:1}(b), \ref{fig:5}(b), and \ref{fig:8}(b).

Table~\ref{tab:d6h} gives the distributions for benzene and coronene, containing 30 and 102 internal vibrational coordinates. Its entries count coordinates, including degeneracy; they are not irreducible-representation multiplicities. Methane has $\Gamma_{\mathrm{vib}}=A_1\oplus E\oplus2T_2$, summarized as multiplicities in Table~\ref{tab:td}.
\begin{inplacetable}
 \caption{Vibrational-coordinate distribution of benzene and coronene ($D_{6h}$). The molecular columns include degeneracy. Divide by the representation dimension to obtain multiplicities. ``Yes'' indicates Raman activity.}
 \label{tab:d6h}
 \begin{ruledtabular}
 \begin{tabular}{lrrrrl}
 Irrep & Dim. & Benzene & Coronene & IR & Raman\\
 $A_{1g}$&1&2&6&---&Yes\\
 $A_{2g}$&1&1&5&---&---\\
 $B_{2g}$&1&2&4&---&---\\
 $B_{1g}$&1&0&2&---&---\\
 $E_{1g}$&2&2&10&---&Yes\\
 $E_{2g}$&2&8&24&---&Yes\\
 $A_{2u}$&1&1&3&$z$&---\\
 $E_{1u}$&2&6&22&$x,y$&---\\
 $B_{1u}$&1&2&6&---&---\\
 $B_{2u}$&1&2&6&---&---\\
 $E_{2u}$&2&4&12&---&---\\
 $A_{1u}$&1&0&2&---&---\\
 \end{tabular}
 \end{ruledtabular}
\end{inplacetable}
\begin{inplacetable}
 \caption{Vibrational-mode multiplicities of methane ($T_d$). Multiplying each multiplicity by its dimension gives a total of nine internal coordinates.}
 \label{tab:td}
 \begin{ruledtabular}
 \begin{tabular}{lrrll}
 Irrep & Dimension & Multiplicity & IR & Raman\\
 $A_1$&1&1&---&Yes\\
 $A_2$&1&0&---&---\\
 $E$&2&1&---&Yes\\
 $T_1$&3&0&---&---\\
 $T_2$&3&2&$x,y,z$&Yes\\
 \end{tabular}
 \end{ruledtabular}
\end{inplacetable}

\looseness=-1 Among the $D_{6h}$ channels, only $E_{1u}$ is both degenerate and infrared active for in-plane driving. Benzene has three such pairs, consistent with its three displayed peaks, and coronene has 11, including the strong modes at $35.8$ and $39.0\,\mathrm{THz}$. The analogous methane channel is $T_2$. An $A_{2u}$ mode is infrared active but one dimensional; an $E_{2g}$ mode is degenerate but electric-dipole forbidden. Even within an allowed degenerate channel, local atomic displacements must be checked: the high-frequency stretching examples in benzene and methane are reported to remain nearly collinear, whereas the lower-frequency examples support elliptical motion.

\looseness=-1 Degeneracy also distinguishes two- and three-component dynamical spaces. $D_{6h}$ has no three-dimensional irreducible representation; the $E_{1u}$ motions considered here lie in the molecular plane and produce perpendicular moments. Methane's $T_2$ space admits three components, with the transverse dynamics discussed in Sec.~\ref{sec:methane} requiring an explicit coupling mechanism. Within a two-dimensional manifold, circular combinations such as $(Q_x+iQ_y)/\sqrt2$ provide a convenient basis for angular momentum and rotating motion. Thus, infrared selection rules identify accessible channels, degeneracy supplies candidate subspaces, and local displacement geometry determines whether their coherent superpositions generate the circulating trajectories relevant to the vibrational IFE.

\end{document}